\documentclass{iopjournal}

\usepackage{amsmath}
\usepackage{amssymb}
\usepackage{booktabs}
\usepackage{multirow}
\usepackage{tikz}
\usepackage{pgfplots}
\pgfplotsset{compat=1.18}
\usepackage[compress]{cite}
\usepackage{microtype}
\usepackage{ragged2e}
\usepackage{graphicx}

\hypersetup{colorlinks=true,linkcolor=blue,filecolor=magenta,citecolor=blue,urlcolor=blue}

\AtBeginDocument{\justifying\setlength{\RaggedRightParindent}{\parindent}}
\newcommand{\doi}[1]{\href{https://doi.org/#1}{\textcolor{blue}{doi:\detokenize{#1}}}}

\begin{document}

\articletype{Paper}

\title{Bound state solutions of the Schrödinger equation for the atomic systems interacting with the radial screened Coulomb potential: analytical approximation methods}

\author{Khaled FZ$^1$\orcid{0009-0005-2853-2265}, Moumni M$^{1,2}$\orcid{0000-0002-8096-6280} and Falek M$^{2,3}$\orcid{0000-0002-0466-9559}}

\affil{$^1$LPRIM, Department of Physics, University of Batna I, Batna, 05000, Algeria}
\affil{$^2$LPPNNM, Department of Matter Sciences, University of Biskra, Biskra, 07000, Algeria}
\affil{$^3$Faculty of Technology, University of Khenchela, Khenchela, 40000, Algeria}

\email{fatmazohra.khaled@univ-batna.dz and m.moumni@univ-batna.dz}


\begin{abstract}
\justifying
We investigate the bound state properties of the hydrogen-like atoms in the radial screened Coulomb potential (RSCP). using three complementary analytical approaches - expectation values with Coulomb and Kratzer reference states, variational optimization with a scaled Kratzer basis, and the Hellmann-Feynman theorem - we derive approximate energy eigenvalues as function of the screening parameter $c$. Benchmarked against high-precision generalized pseudospectral data, the expectation value-approach with the Kratzer basis achieves relative errors of $0.63\%$ for the first ten s-states at $c=0.1$, while the variational method improves this further. The formalism extends naturally to Positronium, demonstrating its generality for arbitrary reduced-mass systems. The complementary biases of the methods provide robust error estimation for plasma-embedded atoms.
\end{abstract}

%=============================================================================
\section{Introduction}
\label{sec:intro}
%=============================================================================

Understanding how atomic systems respond to environmental interactions remains a central issue in quantum physics. When atoms or ions are embedded in plasma or condensed media, their energy spectra and related observables depart from free space values due to many body screening and collective effects. Consequently, the development of accurate potential models is imperative for quantitative predictions on such systems \cite{Jaskolski1996,Sil2005,Paul2009}.
A fundamental model in this context is the screened Coulomb potential (SCP), or Yukawa potential \cite{Yukawa1935}, which describes the interactions of a test charge in weakly coupled plasmas
\begin{equation}
V_Y(r) = - \frac{Ze^2}{r} e^{-\mu r}, \label{eq:scp}
\end{equation}
where  $Z$ denotes the nuclear charge and  $\mu$ represents the screening parameter, which is inversely related to the Debye screening length$\lambda_D = 1/\mu$. The Yukawa form appears across various physical situations from nuclear interactions \cite{Latter1955,Sternheimer1971} to Thomas-Fermi and Debye-Hückel descriptions of screened charges in condensed matter and plasma  \cite{Weisheit1989,Salzmann1998,Murillo1998}. It remains a benchmark for studies of spectral properties under screening \cite{Bielinska2001,Qi2009,Lin2010,Ghoshal2010,Edwards2017,Napsuciale2021,Ghoshal2011,Ghoshal2009}.

To capture more complex screening features, refined forms of the SCP have been proposed. Among these is the exponential cosine screened Coulomb potential (ECSCP), which describes oscillatory screening in strongly coupled plasmas \cite{Takimoto1959,Ikhdair1993,Nasser2011,Lai2013,Bahar2014,Falaye2015,He2023}
\begin{equation}
V_{\text{ECSCP}}(r) = - \frac{Ze^2}{r} e^{-\mu r} \cos(\mu r).  \label{eq:ecscp}
\end{equation}
This model has found applications in the study of quantum dots \cite{Marklund2006,Shukla2010}and semiconductors. More recently, generalized screened potentials incorporating additional polynomial or trigonometric modulations, designated as the more generalized exponential screened Coulomb potential (MGESCP), have been introduced to better describe dense and strongly coupled plasmas \cite{Soylu2012,Jiao2014}. These models have been employed in high precision studies of screening, resonance structure, and spectral shifts\cite{Paul2008,Ghoshal2009,Jiao2014,Rej2017}. 

Recently, Stachura proposed the radial screened Coulomb potential (RSCP) \cite{Stachura2021}, with regular behavior at the origin, screened at short distances and unscreened Coulomb tail at large distances
\begin{equation}
V_{\text{RSCP}}(r) = - \frac{Ze^2}{r} e^{-c/r},  \label{eq:rscp}
\end{equation}
where$c > 0$ is the radial screening parameter. Unlike the Yukawa potential, which exhibits a Coulombic singularity at $r \to 0$, the RSCP is nonsingular everywhere
\begin{equation}
\lim_{r\to 0} V_{\text{RSCP}}(r) = 0, \lim_{r\to\infty} V_{\text{RSCP}}(r) = - \frac{Ze^2}{r}.  \label{eq:limrscp}
\end{equation}
This behaviour makes the RSCP particularly suitable for modelling short-range screening effects while maintaining proper long-range Coulombic asymptotic. Notably, the potential reaches its minimum value$V(c) = -(Ze^2/c)e^{-1}$ at $r=c$, providing a natural length scale for screening.

The RSCP arises naturally in several physical contexts where screening effects depend on proximity to the nucleus rather than on distance from it. From a microscopic perspective, the exponential factor $e^{-c/r}$ can be understood as arising from finite nuclear size effects or from electron cloud distributions that effectively shield the nuclear charge at small distances\cite{Friar1975,Sick2003}. In the context of muonic atoms, where the muon orbits much closer to the nucleus than electrons, similar radial screening effects emerge from the nuclear charge distribution \cite{Pohl2010,Antognini2013}. The RSCP provides a phenomenological model that captures these effects through a single screening parameter. The key distinction between the RSCP and the conventional Yukawa potential lies in their screening regimes. The Yukawa potential $V_Y(r) = -(Ze^2/r)e^{-\mu r}$ screens the Coulomb interaction at large distances ($r \gg 1/\mu$), corresponding to Debye-Hückel screening in weakly coupled plasmas where collective effects dominate beyond the Debye length$\lambda_D = 1/\mu$ \cite{Debye1923,Murillo1998}. In contrast, the RSCP screens the interaction at small distances ($r \lesssim c$), making it appropriate for situations where core polarisation or inner shell screening modifies the effective nuclear potential experienced by valence electrons  \cite{Mitroy2010}. The two potentials thus describe complementary physical regimes: the Yukawa form applies when external collective screening dominates, whilst the RSCP applies when internal atomic or nuclear structure effects are significant.

 Experimentally, radial type screening effects have been inferred from precision spectroscopy of exotic atoms. Measurements of the Lamb shift in muonic hydrogen\cite{Pohl2010} and muonic deuterium\cite{Pohl2016} revealed discrepancies with theoretical predictions based on point nucleus Coulomb potentials, suggesting finite size and screening corrections that can be modelled phenomenologically using potentials of the RSCP type. Similarly, electron scattering experiments probing nuclear charge distributions \cite{DeVries1987,Angeli2013} provide data that constrain the radial dependence of effective screening. In dense astrophysical plasmas, such as white dwarf interiors and neutron star envelopes, pressure ionisation and strong correlations lead to complex screening behaviours that may incorporate both Yukawa type and radial type components \cite{Potekhin1999,Chabrier2002}. From a theoretical standpoint, the RSCP has been employed as a test potential for density functional uniqueness theorems \cite{Stachura2021} and as a model system for developing and benchmarking computational methods  \cite{Stachura2021,Xu2023}. Its mathematical tractability arising from the separability of the radial  Schrödinger equation and the availability of analytical reference solutions makes it an ideal testing ground for approximation techniques that can subsequently be applied to more complex screened potentials.

The accurate determination of atomic energy levels in plasma environments has direct applications in plasma diagnostics and spectroscopy. Line shifts and broadening in emission and absorption spectra provide information about plasma density, temperature, and composition \cite{Griem1997,Fujimoto2004}. Screened Coulomb models, including the RSCP, enable the calculation of these spectral shifts as functions of plasma parameters, thereby connecting theoretical predictions to experimental observables \cite{Salzmann1998,Rosmej2021}. In inertial confinement fusion (ICF) experiments, X-ray spectroscopy of highly charged ions serves as a primary diagnostic tool \cite{Lindl2004,Hurricane2014}.The interpretation of spectral data requires accurate atomic structure calculations that account for plasma screening effects. Similarly, in magnetic confinement devices such as tokamaks, impurity line radiation provides crucial information about plasma conditions and transport \cite{Behringer1991,Isler1994}The RSCP, with its analytical tractability and physical relevance for short-range screening, offers a complementary approach to standard Debye-Hückel models for these applications.

The choice of the RSCP over other screened potentials is motivated by several considerations. First, its nonsingular behaviour at the origin avoids the need for regularisation procedures required by singular potentials. Second, its correct long-range Coulombic asymptotic ensures proper Rydberg state behaviour. Third, the single-parameter form enables systematic studies of screening effects whilst maintaining analytical tractability. Finally, the RSCP interpolates between the unscreened Coulomb limit  ($c \to 0$) and complete screening ($c \to \infty$) in a physically reasonable manner.

Building on these developments, the present work investigates the bound state properties of the hydrogen atom embedded in plasma under the influence of the RSCP. We employ three complementary analytical approaches.  First, using trial wave functions motivated by the Taylor expansion of
the potential (Coulomb-like and Kratzer-like), we derive closed form analytical energy expressions involving modified Bessel functions. By computing $\langle \Psi | \hat{H}_{\text{RSCP}} | \Psi \rangle$ with these reference eigenfunctions, our approach incorporates screening effects directly into the energy estimate. Second, using a scaled Kratzer basis function with an adjustable parameter that optimises the spatial extent of the wave
function, we achieve systematic improvement over the fixed reference methods through variational optimisation. Third, using the Hellmann-Feynman theorem relating energy derivatives to expectation values, we provide an alternative route to the energies via integration from the unscreened limit, developing analytical formulae for both Coulomb and Kratzer bases, including the effect of implicit  $c$ dependence of the effective angular momentum parameter. The obtained results are benchmarked against available reference values from Stachura and Hancock  \cite{Stachura2021} and the high precision GPS calculations of Xu  \textit{et al.} \cite{Xu2023,Xu2024}, and the validity and range of applicability of all approaches are critically discussed. We also extend the formalism to Positronium, demonstrating the generality of the approach for systems with arbitrary reduced mass. While high-precision numeriacl methods such as the GPS approach provide benchmark-quality data, closed-form analytical expressions offer complementary advantages: they enable rapid parametric scaling studies, prvide direct insight into the functional dependence of the screening parameter, and serve as efficient tools for plasma diagnostics where repeated calculations are required.

The general formula of energy are valid for arbitrary quantum numbers $n$ and $\ell$, providing a unified analytical framework applicable to all bound states for every finite value of c.

The paper is organised as follows. Section~\ref{sec:ansatz} presents the expectation value formulation using Coulomb-like and Kratzer-like reference functions. Section~\ref{sec:variational} develops the variational method with a scaled Kratzer basis. Section~\ref{sec:hellmann_feynman} presents the Hellmann-Feynman approach with both basis types. Section~\ref{sec:results} provides a comprehensive comparison of all methods with reference values and error analysis. Section~\ref{sec:comparison_xu} presents a detailed comparison with the asymptotic approximation of Xu \textit{et al.} \cite{Xu2024}. Section~\ref{sec:positronium} extends the formalism to Positronium. Section~\ref{sec:conclusions} summarises the main conclusions and perspectives. The appendices provide detailed derivations: Appendix~\ref{app:A} for the Coulomb basis energy formula, for the Kratzer basis too and for variational ones using unified expressions, and Appendix~\ref{app:B} for the Hellmann-Feynman formula with Kratzer basis.
%=============================================================================
\section{Hamiltonian Expectation-Value Approach}
\label{sec:ansatz}
%=============================================================================

We work in Hartree atomic units where $\hbar = m_e = e^2 = 4\pi\epsilon_0 = 1$. For the hydrogen atom, the reduced mass is $\mu_{\mathrm{H}} \approx 1$ (in the infinite nuclear mass approximation), the Bohr radius is $a_0 = 1$~a.u and the ground state energy is $E_{1s}^{(0)} = -1/2$~Ha. Throughout this paper, $Z = 1$ for hydrogen, and the screening parameter $c$ is measured in atomic units.

The radial Schr\"{o}dinger equation for the hydrogen atom in the RSCP reads
\begin{equation}
    \left[-\frac{1}{2}\frac{d^2}{dr^2} - \frac{1}{r}\frac{d}{dr}+ \frac{l(l+1)}{2r^2} - \frac{e^{-c/r}}{r}\right]R_{n,l}(r) = E_{n,l}\,R_{n,l}(r),
    \label{eq:schrodinger_R}
\end{equation}

The potential in eq.~(\ref{eq:rscp}) can be expanded as a power series in $c/r$
\begin{equation}
    V_{\mathrm{RSCP}}(r) = -\frac{1}{r}\sum_{j=0}^{\infty}\frac{(-1)^{j}}{j!}\left(\frac{c}{r}\right)^{j} = -\frac{1}{r} + \frac{c}{r^2} - \frac{c^2}{2r^3} +  \cdots
    \label{eq:taylor}
\end{equation}
This expansion motivates our choice of reference functions: truncation at different orders yields potentials with known analytical solutions, which provide suitable basis functions for computing approximate energies. Crucially, this expansion serves only as a heuristic guide for constructing auxiliary reference Hamiltonians. In the subsequent expectation-value calculations, we do not truncate the physical potential; the full, exact RSCP $-e^{c/r}/r$ is retained throughout. The approximation enters exclusively through the choice of trial wavefunctions, not through any truncation of the interaction itself. Our approach consists in decomposing the RSCP Hamiltonian by adding and subtracting a reference potential whose eigenfunctions are known analytically
\begin{equation}
    H = H_{\mathrm{ref}} + \Delta V,
    \label{eq:hamiltonian_decomp_general}
\end{equation}
where $H_{\mathrm{ref}}$ possesses known eigenfunctions $\{P_{n}^{\mathrm{ref}}\}$ with eigenvalues $\{E_{n}^{\mathrm{ref}}\}$, and $\Delta V = V_{\mathrm{RSCP}} - V_{\mathrm{ref}}$ is the residual potential. We compute the expectation value of the full RSCP Hamiltonian using the reference eigenfunctions as trial wave functions. The energy expectation value reads
\begin{equation}
    \langle E \rangle = E_{\mathrm{ref}} + \langle \Delta V \rangle_{\mathrm{ref}},
    \label{eq:energy_expectation_general}
\end{equation}
where all expectation values are evaluated with the reference eigenfunctions. We consider two reference potentials corresponding to successive truncation orders of the Taylor expansion~(\ref{eq:taylor}).

\subsection{Case~I: Coulomb Reference Functions ($j = 0$ Truncation)}
\label{subsec:coulomb}

Truncating the expansion~(\ref{eq:taylor}) at $j=0$ yields the pure Coulomb potential $V^{(0)}(r) = -1/r$, which corresponds to the unscreened limit $c\to 0$. The full reduced Hamiltonian is then decomposed as
\begin{equation}
    H = \underbrace{\left[-\frac{1}{2}\frac{d^{2}}{dr^{2}} + \frac{l(l+1)}{2r^{2}} - \frac{1}{r}\right]}_{H_{\mathrm{Coulomb}}} + \underbrace{\frac{1-e^{-c/r}}{r}}_{\Delta V^{(0)}},
    \label{eq:hamiltonian_coulomb}
\end{equation}
where the perturbation $\Delta V^{(0)} = (1 - e^{-c/r})/r$ is manifestly positive for all $c > 0$ and $r > 0$ reflecting the reduction in the attractive potential caused by screening. The reference Coulomb Hamiltonian $H_{\mathrm{Coulomb}}$ for hydrogen has the eigenvalues $E_n^{(0)} = -1/2n^2$,and the normalised reduced radial eigenfunctions of this reduced form of the Hamiltonian are \cite{Schiff1968,Bethe2013}
\begin{equation}
    P_{n,l}^{\mathrm{C}}(r) = \frac{1}{n}\sqrt{\frac{(n\!-\!l\!-\!1)!}{(n\!+\!l)!}}\;\left(\frac{2r}{n}\right)^{l+1} e^{-r/n}\, L_{n-l-1}^{2l+1}\!\!\left(\frac{2r}{n}\right),
    \label{eq:hydrogen_wf}
\end{equation}
$L_{k}^{\alpha}$ denotes the associated Laguerre polynomial. For these solutions, the standard mean values are
\begin{equation}
    \left\langle \frac{1}{r}\right\rangle_{n}^{\!\mathrm{C}} = \frac{1}{n^{2}}, \qquad
    \left\langle \frac{1}{r^2}\right\rangle_{n,l}^{\!\mathrm{C}} = \frac{1}{n^{3}(l+\frac{1}{2})}.
    \label{eq:coulomb_expectation}
\end{equation}

Using eq.~(\ref{eq:energy_expectation_general}), the approximate RSCP energy with the Coulomb reference reads
\begin{equation}
    \langle E_{n,l}\rangle^{\mathrm{C}}
    = E_{n}^{(0)} + \left\langle\frac{1}{r}\right\rangle_{n}^{\!\mathrm{C}}
      - \left\langle\frac{e^{-c/r}}{r}\right\rangle_{n,l}^{\!\mathrm{C}}
    = \frac{1}{2n^{2}}
      - \left\langle\frac{e^{-c/r}}{r}\right\rangle_{n,l}^{\!\mathrm{C}}.
    \label{eq:energy_coulomb_decomp}
\end{equation}
The key quantity is the screened expectation value $\langle e^{-c/r}/r\rangle$. After introducing $\rho = 2r/n$ expanding the Laguerre polynomials, each resulting integral is computed with the master formulas \cite{Gradshteyn2007}
\begin{equation}
    \int_{0}^{\infty}\!x^{\nu-1}\,e^{-\beta/x\,-\,\gamma x}\,dx
    = 2\!\left(\frac{\beta}{\gamma}\right)^{\!\nu/2}\!K_{\nu}\!\left(2\sqrt{\beta\gamma}\right),
    \label{eq:bessel_integral}
\end{equation}
where $K_{\nu}$ is the modified Bessel function of the second kind (MacDonald function). After performing the algebra (see Appendix~\ref{app:A}), the general result for an arbitrary state $(n,l)$ is
\begin{equation}
    \langle E_{n,l} \rangle^{\mathrm{C}} = \frac{1}{2n^{2}} - 2 \frac{(n-l-1)!(n+l)!}{n^{2}}\;\eta_{n,l}^{\mathrm{C}},
    \label{eq:energy_general_C}
\end{equation}
\begin{equation}
    \eta_{n,l}^{\mathrm{C}} = \sum_{m=0}^{n-l-1} \sum_{m'=0}^{n-l-1} \frac{(-1)^{m+m'} \left(\frac{2c}{n}\right)^{(2l+2+m+m')/2} K_{2l+2+m+m'}\left(2\sqrt{\frac{2c}{n}}\right)}{m!\,m'!\,(n\!-\!l\!-\!1\!-\!m)!\,(n\!-\!l\!-\!1\!-\!m')!\,(2l\!+\!1\!+\!m)!\,(2l\!+\!1\!+\!m')!}.
    \label{eq:eta_general_C}
\end{equation}
When the double sums are evaluated explicitly for $s$-states ($l=0$), the result can be expressed in closed form involving only $K_{2}$ and $K_{3}$ Bessel functions with a common argument $z_{n} = 2\sqrt{2c/n}$, multiplied by polynomial in $c$. For the ground state ($1s$, $n=1$), we have (Appendix~\ref{app:A})
\begin{equation}
    \langle E_{1s}\rangle^{\mathrm{C}} = \frac{1}{2} - 4c\,K_{2}(2\sqrt{2c}\,).
    \label{eq:E1s_C}
\end{equation}

For the first excited state ($2s$, $n=2$), with $z = 2\sqrt{c}$
\begin{equation}
    \langle E_{2s}\rangle^{\mathrm{C}} = \frac{1}{8} - \frac{c}{4}\Big[(4+c)\,K_{2}(2\sqrt{c}\,) - \sqrt{c}\,K_{3}(2\sqrt{c}\,)\Big].
    \label{eq:E2s_C}
\end{equation}

For the second excited state ($3s$, $n=3$), with $z = 2\sqrt{2c/3}$
\begin{equation}
 \langle E_{3s}\rangle^{\mathrm{C}}= \frac{1}{18} - \frac{4 c}{729}\left[(3+c)(27+c)\,K_{2}\left(2\sqrt{\frac{2c}{3}}\right) - \sqrt{6c}\,(18+c) K_{3}\left(2\sqrt{\frac{2c}{3}}\right)\right].
    \label{eq:E3s_C}
\end{equation}

As $c \to 0$, the asymptotic expansion $K_{\nu}(z) \sim \frac{\Gamma(\nu)}{2}(2/z)^{\nu}$ ensures that $\langle e^{-c/r}/r\rangle \to \langle 1/r\rangle = 1/n^{2}$, so that each energy correctly reduces to the unscreened Coulomb value
\begin{equation}
    \lim_{c\to 0}\langle E_{n,0}\rangle^{\mathrm{C}} = \frac{1}{2n^{2}} - \frac{1}{n^{2}} = -\frac{1}{2n^{2}} = E_{n}^{(0)}.
    \label{eq:coulomb_limit_C}
\end{equation}

%-----------------------------------------------------------------------
\subsection{Case~II: Kratzer Reference Functions ($j = 1$ Truncation)}
\label{subsec:kratzer}

Including the first-order correction in expansion~(\ref{eq:taylor}) yields the modified Coulomb potential $V^{(1)}(r) = -1/r + c/r^{2}$,whose additional repulsive $1/r^{2}$ term partially accounts for the inner screening of the RSCP. The Hamiltonian is decomposed as
\begin{equation}
    H = \underbrace{\left[-\frac{1}{2}\frac{d^{2}}{dr^{2}} + \frac{\nu(\nu+1)}{2r^{2}} - \frac{1}{r}+\frac{c}{r^{2}}\right]}_{H_{\mathrm{Kratzer}}}
    + \underbrace{\frac{1}{r} - \frac{c}{r^{2}} - \frac{e^{-c/r}}{r}}_{\Delta V^{(1)}},
    \label{eq:hamiltonian_kratzer}
\end{equation}
where the effective angular momentum quantum number is defined by
\begin{equation}
     \nu_{l}= -\tfrac{1}{2} + \sqrt{\left(l+\tfrac{1}{2}\right)^{2}+2c\,},
    \label{eq:nu_def}
\end{equation}
satisfying $\nu(\nu+1) = l(l+1)+2c$, with $\nu_{l}\to l$ as $c\to 0$.

The Kratzer Hamiltonian is formally identical to the Coulomb Hamiltonian with $l$ replaced by $\nu$. For hydrogen, the eigenvalues are $E_{n_{r},l}^{\mathrm{Kr}} = -12N_{\nu}^{2}$, where $n_{r} = 0,1,2,\ldots$ is the radial quantum number and the effective principal quantum number is $N_{\nu} = n_{r} + \nu_{l} + 1$. The normalised reduced radial eigenfunctions are
\begin{equation}
    P_{n_{r},l}^{\mathrm{Kr}}(r)
    = \frac{1}{N_{\nu}}\sqrt{\frac{\Gamma(n_{r}+1)}{\Gamma(n_{r}+2\nu_{l}+2)}}
      \left(\frac{2r}{N_{\nu}}\right)^{\!\nu_{l}+1}
      e^{-r/N_{\nu}}\,
      L_{n_{r}}^{2\nu_{l}+1}\!\!\left(\frac{2r}{N_{\nu}}\right),
    \label{eq:kratzer_wf}
\end{equation}
where for non-integer $\nu_{l}$, the Laguerre polynomials $L_{n_{r}}^{2\nu_{l}+1}$ are defined through the generalised series representation and factorials are replaced by Gamma functions. The standard expectation values with Kratzer eigenfunctions are
\begin{align}
    \left\langle\frac{1}{r}\right\rangle_{n_{r}}^{\!\mathrm{Kr}} = \frac{1}{N_{\nu}^{2}}
   , \quad  \left\langle\frac{1}{r^{2}}\right\rangle_{n_{r}}^{\!\mathrm{Kr}} = \frac{1}{N_{\nu}^{3}\,(\nu_{l}+\tfrac{1}{2})}.
    \label{eq:kratzer_1r2}
\end{align}

Using eq.~(\ref{eq:energy_expectation_general}) with eq.~(\ref{eq:hamiltonian_kratzer}) and the Kratzer expectations values, the energy expression is
\begin{equation}
\langle E_{n_{r},l}\rangle^{\mathrm{Kr}}  = \frac{1}{2N_{\nu}^{2}} - \frac{c}{N_{\nu}^{3}\sqrt{(l+\tfrac{1}{2})^{2}+2c}} - \left\langle\frac{e^{-c/r}}{r}\right\rangle_{n_{r},l}^{\!\mathrm{Kr}}.
    \label{eq:energy_final_Kr}
\end{equation}

The expectation value $\langle e^{-c/r}/r\rangle^{\mathrm{Kr}}$ is evaluated by substituting $\rho = 2r/N_{\nu}$, expanding the generalised Laguerre polynomial, and applying the master integral formula~(\ref{eq:bessel_integral}).

For $s$-states we set $l = 0$, giving $\nu_{l=0} = -1/2+\sqrt{2c+1/4}$ and $N_{\nu} = n_{r}+\nu_{0}+1$ (We use the notation $\nu \equiv \nu_{0}$ for brevity). Since $\nu$ is generally non-integer for $c > 0$, all factorials involving $\nu$ or $N_{\nu}$ must be replaced by Gamma functions using the identity $n! = \Gamma(n+1)$.

For arbitrary angular momentum $l$, the screened expectation value is computed using
\begin{equation}
    \left\langle\frac{e^{-c/r}}{r}\right\rangle_{n_{r},l}^{\!\mathrm{Kr}}
    = \frac{\Gamma(n_r+1)}{N_\nu^{2} \Gamma(n_r+2\nu_{l}+2)} \mathcal{I}_{n_{r},l}^{\mathrm{RSCP}},
    \label{eq:rscp_expectation_general}
\end{equation}
with $\mathcal{I}_{n_{r},l}^{\mathrm{RSCP}}$ defined by a double sum over modified Bessel functions as detailed in Appendix~\ref{app:A}. In the limit $c\to 0$ we have $\nu_{l}\to l$, $N_{\nu}\to n_{r}+l+1 = n$, and the Kratzer eigenfunctions reduce to the standard Coulomb eigenfunctions. The expectation values then recover their Coulomb counterparts, and  eq.~(\ref{eq:energy_final_Kr}) reduces to the Coulomb-reference result~(\ref{eq:energy_coulomb_decomp}), confirming the consistency of the approach.

%-----------------------------------------------------------------------
\subsection{Physical Interpretation of Biases}
\label{subsec:physical}

The RSCP at small $r$ behaves as
\begin{equation}
    V_{\mathrm{RSCP}}(r) \approx -\frac{1}{r} + \frac{c}{r^{2}} - \frac{c^{2}}{2r^{3}} + \cdots
    \label{eq:rscp_expansion_phys}
\end{equation}

The Kratzer reference incorporates the leading $c/r^{2}$ correction, which modifies the effective centrifugal barrier from $l(l+1)/r^{2}$ to $\nu(\nu+1)/r^{2}$ with $\nu > l$ for $c > 0$. This pushes the wave function outward, providing a better zeroth order approximation than the Coulomb reference.

The two reference methods exhibit complementary biases
\begin{itemize}
    \item \textbf{Coulomb reference:} The Coulomb wave function samples the small $r$ region where the perturbation $\Delta V^{(0)} = (1-e^{-c/r})/r$ is large. Neglecting the repulsive $c/r^2$  term leads to an underestimate of binding (energies too close to zero), particularly for the ground state at moderate and large~$c$.

    \item \textbf{Kratzer reference:} The Kratzer wave function accounts for the $c/r^2$ repulsion but does not fully capture the exponential suppression  $e^{-c/r}$ at small $r$. The Kratzer eigenstates remain slightly more compact than the true RSCP eigenstates, leading to an overestimate of binding (energies too negative) for moderate and large~$c$.
\end{itemize}

The physical reason for the superior performance of the Kratzer reference can be understood as follows. The RSCP modifies the effective potential experienced by the electron primarily through the $c/r^2$ term in the Taylor expansion. This term has the same radial dependence as the centrifugal barrier, so it can be absorbed exactly into an effective angular momentum $\nu > l$. The Kratzer eigenfunctions, being exact solutions of the Coulomb problem with this modified centrifugal barrier, automatically incorporate the dominant effect of screening on the wave function shape. In contrast, the Coulomb eigenfunctions are optimised for the unscreened case and must account for the entire screening effect through the perturbation term, leading to larger errors.

It is worth emphasising that our approach is fundamentally distinct from standard perturbation theory. In conventional perturbation method, the potential itself is expanded and truncated, and the corrections are computed order by order. here by contrast, we evaluate the expectation value of the \textbf{\textit{exact, full RSCP Hamiltonian}} $\langle\Psi_{\mathrm{ref}}|\hat{H}_{\mathrm{RSCP}}|\Psi_{\mathrm{ref}}\rangle$. The approximation is therefore confined entirely to the trial wavefunctions, while the potential retains its exact nonsingular form and correct long-range Coulomb tail. This distinction is crucial: the perturbative approach computes $\langle\Psi|\hat{H}_{\mathrm{approx}}|\Psi\rangle$ where $\hat{H}_{\mathrm{approx}}$ retains only a finite number of terms in the Taylor expansion of$e^{-c/r}$, where as our approach uses approximate wave functions but the exact Hamiltonian, ensuring that all orders of the screening are captured in the energy estimate. Although the auxiliary Coulomb and Kratzer reference potentials are singular at the origin, this is an artefact of the basis choice and does not imply a singularity in the physical RSCP. Since the expectation value is evaluated with the regular, finite RSCP, the wavefunction singularities are integrated against a smooth potential and pose no mathematical or physical difficulties.

%=============================================================================
\section{Variational Method with Scaled Kratzer Basis}
\label{sec:variational}
%=============================================================================

The results of Section~\ref{sec:ansatz} demonstrate that the expectation value approach with Coulomb and Kratzer bases exhibit complementary biases: the Coulomb reference underestimates binding (energies too close to zero), while the Kratzer reference overestimates binding (energies too negative) for low lying states at moderate screening. The true RSCP energy lies between these two predictions, suggesting that a systematic improvement can be achieved through variational optimisation. The variational principle states that for any normalised trial wave function $\psi_{\mathrm{trial}}$,the energy expectation value provides an upper bound to the true ground state energy:
\begin{equation}
E_{\mathrm{var}} = \frac{\langle \psi^{\mathrm{trial}} | H | \psi^{\mathrm{trial}} \rangle}{\langle \psi^{\mathrm{trial}} | \psi^{\mathrm{trial}} \rangle} \geq E_{\mathrm{true}},
\label{eq:variational_principle}
\end{equation}
with equality holding only when $\psi_{\mathrm{trial}}$ coincides with the exact eigenfunction. As the expectation-value approach, the variational method uses the \textbf{full, unexpanded RSCP} to evaluate the energy. The following optimisation is therefore performed with the exact potential, not the truncated series.

By introducing one or more adjustable parameters into the trial function and minimising $E_{\mathrm{var}}$ with respect to these parameters, we obtain the best approximation within the chosen functional form. Since the Kratzer reference already incorporates the leading $c/r^{2}$ correction and achieves relative errors below $0.63$\% (Table \ref{tab:rscp_combined_0.1}) for the first ten $s$-states at $c = 0.1$, it provides an excellent starting point for variational refinement. We construct the trial wave function by introducing a positive scaling parameter $\beta$ into the Kratzer eigenfunctions
\begin{equation}
P_{n_{r}}^{\mathrm{trial}}(r;\beta) = \mathcal{N}(\beta) \left(\frac{2\beta r}{N_{\nu}}\right)^{\!\nu+1} e^{-\beta r/N_{\nu}} L_{n_{r}}^{2\nu+1}\!\left(\frac{2\beta r}{N_{\nu}}\right),
\label{eq:trial_wf}
\end{equation}
where $\nu = \nu_{0} = -1/2 + \sqrt{2c+1/4}$ for $s$-states ($l=0$), $N_{\nu} = n_{r} + \nu + 1$, and $\mathcal{N}(\beta)$ is the $\beta$-dependent normalisation constant. The scaling parameter $\beta$ controls the spatial extent of the wave function: $\beta = 1$ recovers the unscaled Kratzer eigenfunction from Section~\ref{subsec:kratzer}; $\beta < 1$ expands the wave function, reducing the kinetic energy but increasing the potential energy expectation value (less binding); $\beta > 1$ contracts the wave function, increasing the kinetic energy but lowering the potential energy expectation value (more binding). These scaled functions provide accurate approximations for excited states within the chosen ansatz, although the strict variational upper bound property for excited states requires orthogonality constraints to lower states.

The variational energy is obtained by evaluating the expectation value of the RSCP Hamiltonian with the trial function~(\ref{eq:trial_wf})
\begin{equation}
E_{\mathrm{var}}(\beta) = \langle T \rangle_{\beta} + \langle V_{\mathrm{eff}} \rangle_{\beta} + \langle V_{\mathrm{RSCP}} \rangle_{\beta},
\label{eq:var_energy}
\end{equation}
Here we decompose the physical kinetic energy operator $\hat{T} = -\frac{1}{2}d^{2}/dr^{2}$ acting on the trial function into a 'reduced kinetic energy' $\langle T\rangle_{\beta}$ and an `effective centrifugal' contribution $\langle V_{\mathrm{eff}}\rangle_{\beta}
= \nu(\nu+1)\langle 1/(2r^{2})\rangle_{\beta}$ arising from the $r^{\nu+1}$ prefactor of the Kratzer wave function.  Their sum equals
$\langle -\frac{1}{2}d^{2}/dr^{2}\rangle_{\beta}$ and does not represent an additional physical potential. All expectation values can be computed analytically using the change of variables $\rho = 2\beta r / N_{\nu}$ and the techniques developed in the Appendices.  To get the normalisation constant,  we compute $\int_0^\infty |P_{n_r}^{\mathrm{trial}}(r;\beta)|^2\,dr$ using the substitution $\rho = 2\beta r/N_\nu$ the standard Laguerre orthogonality relation
\begin{align}
\int_0^\infty |P_{n_r}^{\mathrm{trial}}|^2\,dr &= |\mathcal{N}(\beta)|^2 \left(\frac{N_\nu}{2\beta}\right) \int_0^\infty \rho^{2\nu+2} e^{-\rho} \left[L_{n_r}^{2\nu+1}(\rho)\right]^2 d\rho \nonumber \\
&= |\mathcal{N}(\beta)|^2 \left(\frac{N_\nu}{2\beta}\right) \frac{2 N_{\nu}\Gamma(n_r+2\nu+2)}{n_r!}=1,
\label{eq:norm_check}
\end{align}
\begin{equation}
\implies \mathcal{N}(\beta) = \frac{\sqrt{\beta}}{N_{\nu}} \sqrt{\frac{n_{r}!}{\Gamma(n_{r} + 2\nu + 2)}}.
\label{eq:normalization_constant}
\end{equation}

The kinetic energy and effective centrifugal expectation values scale as $\beta^{2}$
\begin{equation}
\langle T \rangle_{\beta} + \langle V_{\mathrm{eff}} \rangle_{\beta} = \frac{\beta^{2}}{2N_{\nu}^{2}}\left(1 - \frac{2c}{N_{\nu}(\nu+\frac{1}{2})}\right).
\label{eq:kinetic_centrifugal_result}
\end{equation}

The RSCP potential expectation value is obtained from the Kratzer formula of Section~\ref{subsec:kratzer} by the substitution $2c/N_{\nu} \to 2c\beta/N_{\nu}$ and $z = 2\sqrt{2c/N_{\nu}} \to \tilde{z} = 2\sqrt{2c\beta/N_{\nu}}$. For the ground state ($n_{r} = 0$):
\begin{equation}
\langle V_{\mathrm{RSCP}} \rangle_{\beta}^{(1s)} = -\frac{2 \beta (\tilde{c})^{\nu +1}}{N_{\nu}^{2}\Gamma(2\nu+2)} K_{2\nu+2}(\tilde{z}),
\label{eq:rscp_1s_var}
\end{equation}
where $\tilde{c} = 2c\beta/N_{\nu}$, $\tilde{z} = 2\sqrt{\tilde{c}}$, and $N_{\nu} = \nu + 1$.

The total variational energy is
\begin{equation}
E_{\mathrm{var}}(\beta) = \frac{\beta^{2}}{2N_{\nu}^{2}}\left(1 - \frac{2c}{N_{\nu}(\nu+\frac{1}{2})}\right) + \langle V_{\mathrm{RSCP}} \rangle_{\beta},
\label{eq:total_var_energy}
\end{equation}
and the optimal scaling parameter $\beta_{\mathrm{opt}}$ is determined by minimising the variational energy numerically

%=============================================================================
\section{Hellmann-Feynman Approach}
\label{sec:hellmann_feynman}
%=============================================================================

The Hellmann-Feynman theorem \cite{Hellmann1937,Feynman1939} provides an alternative route to the energy eigenvalues by relating the derivative of the energy with respect to a parameter to an expectation value. For the RSCP with screening parameter $c$, the theorem states
\begin{equation}
    \frac{\partial E}{\partial c} = \left\langle \frac{\partial V_{\mathrm{RSCP}}}{\partial c} \right\rangle = \left\langle \frac{e^{-c/r}}{r^2} \right\rangle.
    \label{eq:hellmann_feynman}
\end{equation}
As the expectation-value approach, the Hellmann-Feynman method uses the \textbf{full, unexpanded RSCP} to evaluate the energy. The following integration is therefore performed with the exact potential, not the truncated series.

Since the derivative $\partial V_{\mathrm{RSCP}}/\partial c = e^{-c/r}/r^2 > 0$ for all $r > 0$, it immediately implies that $\partial E/\partial c > 0$. This confirms analytically that the energy increases monotonically with the screening parameter, consistent with our numerical results from the expectation value and variational methods.

Integrating eq.~(\ref{eq:hellmann_feynman}) from $c = 0$ (where the Coulomb energy is known) to a finite value $c$, gives
\begin{equation}
    E(c) = E(0) + \int_0^c \left\langle \frac{e^{-c'/r}}{r^2} \right\rangle dc'.
    \label{eq:HF_integration}
\end{equation}
For exact eigenfunctions, this relation is exact. When approximate wave functions are used, the integration provides an estimate of the energy that depends on the quality of the trial functions employed. We evaluate this approach using both Coulomb and Kratzer eigenfunctions.

\subsection{Hellmann-Feynman with Coulomb Basis (HF-C)}
\label{subsec:HF_coulomb}

Because the Coulomb basis functions are independent of the screening parameter $c$, the Hellmann--Feynman integration in eq.~(\ref{eq:HF_integration}) can be performed inside the expectation value, yielding

\begin{equation}
\int_{0}^{c} \left\langle \frac{e^{-c'/r}}{r^2}\right\rangle^C dc' = \left\langle \frac{1}{r} - \frac{e^{-c/r}}{r}\right\rangle^C.
\label{eq:HFC_eq_CA}
\end{equation}
Therefore, the HF-C method is exactly equivalent to the Coulomb reference method of eq.~(\ref{eq:energy_coulomb_decomp}).

\subsection{Hellmann-Feynman with Kratzer Basis (HF-Kr)}
\label{subsec:HF_kratzer}

Using the Kratzer eigenfunctions~(\ref{eq:kratzer_wf}), the corresponding expectation value for $s$-states is derived in detail in Appendix~\ref{app:B}
\begin{align}
    \left\langle \frac{e^{-c/r}}{r^2} \right\rangle_{n_r,0}^{\mathrm{Kr}} &= \frac{4\,n_r! \Gamma(n_r+2\nu_0+2)}{N_\nu^{3}} \sum_{m=0}^{n_r} \sum_{m'=0}^{n_r} \chi^{n_{r},\nu}_{mm'} \nonumber \\
    &\quad \times \left(\frac{2c}{N_\nu}\right)^{(2\nu_0+1+m+m')/2} K_{2\nu_0+1+m+m'}\left(2\sqrt{\frac{2c}{N_\nu}}\right),
    \label{eq:HF_expectation_general}
\end{align}
where the coefficients $\chi^{n_{r},\nu}_{mm'}$ are defined in \ref{eq:B9} and $\nu_0$ means $\nu$ for $l=0$.

For the ground state ($n_r = 0$, $N_\nu = \nu_0 + 1 = \tfrac{1}{2} + \sqrt{\tfrac{1}{4} + 2c}$):
\begin{equation}
 \left\langle \frac{e^{-c/r}}{r^2} \right\rangle_{1s}^{\mathrm{Kr}} = \frac{4\,(2c/N_\nu)^{\nu_0+1/2}}{N_\nu^{3}\,\Gamma(2\nu_0+2)}\, K_{2\nu_0+1}\left(2\sqrt{\frac{2c}{N_\nu}}\right),
    \label{eq:HF_1s}
\end{equation}

The energy for each state is obtained by integrating the corresponding expectation value
\begin{equation}
    E_{n_r,0}^{\mathrm{HF-Kr}}(c) = -\frac{1}{2n^2} + \int_0^c \left\langle \frac{e^{-c'/r}}{r^2} \right\rangle_{n_r,0}^{\mathrm{Kr}}(c') \, dc',
    \label{eq:HF_energy_general}
\end{equation}
where the unscreened Coulomb energies are $E_{n}^{(0)} = -1/(2n^2)$. Due to the $c$ dependence of the parameters $\nu_0(c)$ and $N_\nu(c)$ in the integrand, the integration is performed numerically.

As $c \to 0$, the results reduce to the Coulomb limit: $\nu_0 \to 0$, $N_\nu \to n$, and the integrand approaches $\langle 1/r^2 \rangle_{n,0}^{\mathrm{C}} = 1/n^{3}$,ensuring that the integral is well defined at the lower limit. Note that while the Hellmann-Feynman theorem strictly requires exact eigenstates, using approximate trial functions yields consistent estimates within the approximation scheme, with the quality of results depending on how well the trial functions approximate the true eigenstates.

%=============================================================================
\section{Results and Discussion}
\label{sec:results}
%=============================================================================
\subsection{Reference Values}
For the validation of our results, we use values from Xu \textit{et al.}~\cite{Xu2023} for the hydrogen atom at $c = 0.1$where they used a high precision GPS method, along with earlier results from Stachura and Hancock~\cite{Stachura2021} and the Hellmann-Feynman based values from Xu \textit{et al.}~\cite{Xu2024}. Table \ref{tab:rscp_combined_0.1} gives our values computed from all methods presented in this work compared to reference values when the parameter $c=0.1$, and also the relative errors of our values compared to the results from GPS in~\cite{Xu2023}. The GPS values represent the most accurate available results and will be used as the primary benchmark. We note that the H-F values from Xu \textit{et al.} ~\cite{Xu2024} systematically underestimate binding (20.9\% error for 1s), while the S-H values~\cite{Stachura2021} overestimate binding (12.5\% error for 1s).
%=============================================================================
\begin{table}[h]
\centering
\caption{Energy eigenvalues $E$ (in Ha) for the first ten $s$ states of the hydrogen atom in the RSCP at $c=0.1$. Relative deviations from the GPS results, $\delta = |(E-E_{\rm GPS})/E_{\rm GPS}|\times 100$, are given in parentheses.}
\label{tab:rscp_combined_0.1}
\resizebox{\textwidth}{!}{%
\begin{tabular}{lccccccc}
\toprule
State & Coulomb ref. & HF-Kr & Kratzer ref. & Variational & GPS~\cite{Xu2023} & Xu H-F (2024)~\cite{Xu2024} & S-H (2021)~\cite{Stachura2021} \\
\midrule
$1s$  & $-3.436212(-1)\,(9.418)$ & $-3.846537(-1)\,(1.399)$ & $-3.769734(-1)\,(0.626)$ & $-3.778187(-1)\,(0.403)$ & $-3.793464(-1)$ & $-3.000000(-1)\,(20.917)$ & $-4.268000(-1)\,(12.509)$ \\
$2s$  & $-1.053204(-1)\,(2.772)$ & $-1.088999(-1)\,(0.533)$ & $-1.079848(-1)\,(0.312)$ & $-1.080506(-1)\,(0.251)$ & $-1.083227(-1)$ & $-1.000000(-1)\,(7.683)$  & $-1.052000(-1)\,(2.883)$ \\
$3s$  & $-4.971678(-2)\,(1.429)$ & $-5.059717(-2)\,(0.317)$ & $-5.033267(-2)\,(0.208)$ & $-5.034677(-2)\,(0.180)$ & $-5.043743(-2)$ & $-4.815000(-2)\,(4.535)$  & $-5.260000(-2)\,(4.288)$ \\
$4s$  & $-2.878560(-2)\,(0.916)$ & $-2.911660(-2)\,(0.223)$ & $-2.900663(-2)\,(0.156)$ & $-2.901129(-2)\,(0.140)$ & $-2.905185(-2)$ & $-2.813000(-2)\,(3.173)$  & $-3.000000(-2)\,(3.264)$ \\
$5s$  & $-1.873795(-2)\,(0.659)$ & $-1.889459(-2)\,(0.171)$ & $-1.883883(-2)\,(0.124)$ & $-1.884079(-2)\,(0.114)$ & $-1.886230(-2)$ & $-1.840000(-2)\,(2.451)$  & $-1.940000(-2)\,(2.851)$ \\
$6s$  & $-1.315844(-2)\,(0.509)$ & $-1.324405(-2)\,(0.139)$ & $-1.321200(-2)\,(0.104)$ & $-1.321296(-2)\,(0.096)$ & $-1.322571(-2)$ & $-1.296000(-2)\,(2.009)$  & $-1.360000(-2)\,(2.830)$ \\
$7s$  & $-9.744060(-3)\,(0.411)$ & $-9.795662(-3)\,(0.116)$ & $-9.775581(-3)\,(0.089)$ & $-9.776110(-3)\,(0.083)$ & $-9.784273(-3)$ & $-9.621000(-3)\,(1.669)$  & $-1.000000(-2)\,(2.205)$ \\
$8s$  & $-7.504307(-3)\,(0.343)$ & $-7.537707(-3)\,(0.100)$ & $-7.524307(-3)\,(0.078)$ & $-7.524620(-3)\,(0.074)$ & $-7.530159(-3)$ & $-7.422000(-3)\,(1.436)$  & $-7.600000(-3)\,(0.927)$ \\
$9s$  & $-5.956379(-3)\,(0.294)$ & $-5.979193(-3)\,(0.088)$ & $-5.969811(-3)\,(0.069)$ & $-5.970010(-3)\,(0.066)$ & $-5.973937(-3)$ & $-5.898000(-3)\,(1.271)$  & $-6.000000(-3)\,(0.436)$ \\
$10s$ & $-4.842196(-3)\,(0.256)$ & $-4.858450(-3)\,(0.078)$ & $-4.851629(-3)\,(0.062)$ & $-4.851760(-3)\,(0.059)$ & $-4.854645(-3)$ & $-4.800000(-3)\,(1.126)$  & $-5.000000(-3)\,(2.994)$ \\
\bottomrule
\end{tabular}%
}
\end{table}
%=============================================================================

Table \ref{tab:comparison_c001_c1_c10} presents results for states with different radial quantum number, demonstrating how the accuracy of our methods depends on $l$. The accuracy of the expectation value approach with Kratzer basis is comparable for all angular momentum values, with relative errors typically below 0.1\%. This uniformity arises because the effective angular momentum $\nu$ correctly captures the modification of the centrifugal barrier for all $l$ via the relation $\nu(\nu+1) = l(l+1) + 2c$. For higher $l$ values, the wave function is pushed further from the origin by the centrifugal barrier, reducing the importance of the short-range screening encoded in the exponential factor $e^{-c/r}$. This explains why the relative errors are often slightly smaller for high $l$-states compared to $s$-states at same principal quantum number.
%=============================================================================
\begin{table}[h]
\centering
\caption{Comparison of energy eigenvalues (in Ha) for selected bound states of the hydrogen atom in the RSCP at $c=0.01$, $1.0$, and $10.0$. The GPS values are taken from Xu \textit{et al.} and used as reference data.}
\label{tab:comparison_c001_c1_c10}
\resizebox{\textwidth}{!}{%
\begin{tabular}{l c c c c c c}
\toprule
\textbf{State} & \boldmath$c$ & \textbf{Coulomb ref.} & \textbf{HF-Kr} & \textbf{Kratzer ref.} & \textbf{Variational} & \textbf{GPS}~\cite{Xu2023} \\
\midrule
\multirow{3}{*}{1s}
 & $0.01$ & $-4.808590(-1)$ & $-4.817046(-1)$ & $-4.816506(-1)$ & $-4.816526(-1)$ & $-4.816627(-1)$ \\
 & $1.0$  & $1.907654(-1)$  & $-2.172052(-1)$ & $-1.413263(-1)$ & $-1.476385(-1)$ & $-1.548831(-1)$ \\
 & $10.0$ & $4.973326(-1)$  & $-1.508729(-1)$ & $-2.371958(-2)$ & $-2.663682(-2)$ & $-2.834503(-2)$ \\
\midrule
\multirow{3}{*}{2s}
 & $0.01$ & $-1.226060(-1)$ & $-1.226789(-1)$ & $-1.226731(-1)$ & $-1.226732(-1)$ & $-1.226746(-1)$ \\
 & $1.0$  & $-3.035335(-2)$ & $-7.344765(-2)$ & $-6.028130(-2)$ & $-6.126718(-2)$ & $-6.381610(-2)$ \\
 & $10.0$ & $9.720836(-2)$  & $-4.672539(-2)$ & $-1.600454(-2)$ & $-1.708517(-2)$ & $-1.859175(-2)$ \\
\midrule
\multirow{3}{*}{3s}
 & $0.01$ & $-5.484613(-2)$ & $-5.486445(-2)$ & $-5.486281(-2)$ & $-5.486283(-2)$ & $-5.486325(-2)$ \\
 & $1.0$  & $-2.698283(-2)$ & $-3.762995(-2)$ & $-3.322701(-2)$ & $-3.351198(-2)$ & $-3.462879(-2)$ \\
 & $10.0$ & $2.315030(-2)$  & $-2.404562(-2)$ & $-1.152232(-2)$ & $-1.202904(-2)$ & $-1.309703(-2)$ \\
\midrule
\multirow{3}{*}{2p}
 & $0.01$ & $-1.241687(-1)$ & $-1.241747(-1)$ & $-1.241747(-1)$ & $-1.241747(-1)$ & $-1.241747(-1)$ \\
 & $1.0$  & $-5.799299(-2)$ & $-8.189449(-2)$ & $-7.942526(-2)$ & $-7.974878(-2)$ & $-7.987591(-2)$ \\
 & $10.0$ & $1.018732(-1)$  & $-4.085989(-2)$ & $-2.125704(-2)$ & $-2.289338(-2)$ & $-2.339193(-2)$ \\
\midrule
\multirow{3}{*}{3p}
 & $0.01$ & $-5.530926(-2)$ & $-5.531061(-2)$ & $-5.531061(-2)$ & $-5.531061(-2)$ & $-5.531061(-2)$ \\
 & $1.0$  & $-3.558994(-2)$ & $-4.133735(-2)$ & $-4.060176(-2)$ & $-4.068229(-2)$ & $-4.075926(-2)$ \\
 & $10.0$ & $2.333126(-2)$  & $-2.251117(-2)$ & $-1.461532(-2)$ & $-1.528715(-2)$ & $-1.580913(-2)$ \\
\midrule
\multirow{3}{*}{4p}
 & $0.01$ & $-3.114609(-2)$ & $-3.114658(-2)$ & $-3.114658(-2)$ & $-3.114658(-2)$ & $-3.114658(-2)$ \\
 & $1.0$  & $-2.281035(-2)$ & $-2.489822(-2)$ & $-2.459002(-2)$ & $-2.461893(-2)$ & $-2.466221(-2)$ \\
 & $10.0$ & $4.098270(-3)$  & $-1.463627(-2)$ & $-1.066263(-2)$ & $-1.099405(-2)$ & $-1.139472(-2)$ \\
\midrule
\multirow{3}{*}{3d}
 & $0.01$ & $-5.540753(-2)$ & $-5.540794(-2)$ & $-5.540794(-2)$ & $-5.540794(-2)$ & $-5.540794(-2)$ \\
 & $1.0$  & $-4.189069(-2)$ & $-4.466369(-2)$ & $-4.450551(-2)$ & $-4.452040(-2)$ & $-4.452387(-2)$ \\
 & $10.0$ & $2.145441(-2)$  & $-2.276958(-2)$ & $-1.765170(-2)$ & $-1.830391(-2)$ & $-1.842539(-2)$ \\
\midrule
\multirow{3}{*}{4d}
 & $0.01$ & $-3.118755(-2)$ & $-3.118770(-2)$ & $-3.118770(-2)$ & $-3.118770(-2)$ & $-3.118770(-2)$ \\
 & $1.0$  & $-2.548105(-2)$ & $-2.646359(-2)$ & $-2.640071(-2)$ & $-2.640585(-2)$ & $-2.640886(-2)$ \\
 & $10.0$ & $2.623937(-3)$  & $-1.500133(-2)$ & $-1.250917(-2)$ & $-1.281047(-2)$ & $-1.295979(-2)$ \\
\midrule
\multirow{3}{*}{5d}
 & $0.01$ & $-1.996803(-2)$ & $-1.996809(-2)$ & $-1.996809(-2)$ & $-1.996809(-2)$ & $-1.996810(-2)$ \\
 & $1.0$  & $-1.704533(-2)$ & $-1.748707(-2)$ & $-1.745625(-2)$ & $-1.745848(-2)$ & $-1.746054(-2)$ \\
 & $10.0$ & $-2.314102(-3)$ & $-1.072121(-2)$ & $-9.326513(-3)$ & $-9.485811(-3)$ & $-9.611312(-3)$ \\
\midrule
\multirow{3}{*}{4f}
 & $0.01$ & $-3.120538(-2)$ & $-3.120544(-2)$ & $-3.120544(-2)$ & $-3.120544(-2)$ & $-3.120544(-2)$ \\
 & $1.0$  & $-2.696571(-2)$ & $-2.749575(-2)$ & $-2.747769(-2)$ & $-2.747882(-2)$ & $-2.747901(-2)$ \\
 & $10.0$ & $-4.759003(-4)$ & $-1.569734(-2)$ & $-1.414005(-2)$ & $-1.436399(-2)$ & $-1.439345(-2)$ \\
\midrule
\multirow{3}{*}{5f}
 & $0.01$ & $-1.997715(-2)$ & $-1.997718(-2)$ & $-1.997718(-2)$ & $-1.997718(-2)$ & $-1.997718(-2)$ \\
 & $1.0$  & $-1.780609(-2)$ & $-1.804074(-2)$ & $-1.803211(-2)$ & $-1.803259(-2)$ & $-1.803279(-2)$ \\
 & $10.0$ & $-4.044581(-3)$ & $-1.120872(-2)$ & $-1.036060(-2)$ & $-1.047496(-2)$ & $-1.051586(-2)$ \\
\midrule
\multirow{3}{*}{6f}
 & $0.01$ & $-1.387567(-2)$ & $-1.387568(-2)$ & $-1.387568(-2)$ & $-1.387568(-2)$ & $-1.387568(-2)$ \\
 & $1.0$  & $-1.261915(-2)$ & $-1.274034(-2)$ & $-1.273560(-2)$ & $-1.273584(-2)$ & $-1.273600(-2)$ \\
 & $10.0$ & $-4.592110(-3)$ & $-8.427236(-3)$ & $-7.916928(-3)$ & $-7.981567(-3)$ & $-8.018741(-3)$ \\
\bottomrule
\end{tabular}%
}
\end{table}
%=============================================================================
\subsection{Comprehensive Comparison of All Methods}
\begin{itemize}
    \item The Coulomb reference and HF-C methods provide a simple approximation but systematically underestimate binding, with errors ranging from 0.26\% to 9.42\%.
    \item The full HF-Kr method achieves further improvement, with errors ranging from 0.08\% to 1.40\%. The $1s$ state remains a numerical outlier, reflecting its high sensitivity to the screening.
    \item The Kratzer reference achieves order of magnitude improvement by incorporating the $c/r^2$ correction directly, reducing errors to below 0.63\% for all ten $s$-states.
    \item The variational method provides the best results, with errors decreasing from 0.40\% for $1s$ to 0.06\% for $10s$.
\end{itemize}

\subsection{Optimal Scaling Parameters}

Table \ref{tab:beta_opt_c01} presents the optimal scaling parameters $\beta_{\mathrm{opt}}$ for the first ten $s$-states at $c=0.1$.

\begin{table}[h]
\centering
\caption{Optimal variational scaling parameter $\beta_{\mathrm{opt}}$ for the first ten $s$-states of H-atom in the RSCP at $c=0.1$.}
\label{tab:beta_opt_c01}
\begin{tabular}{c c c c c c c c c c c}
\toprule
State & $1s$ & $2s$ & $3s$ & $4s$ & $5s$ & $6s$ & $7s$ & $8s$ & $9s$ & $10s$ \\
$\beta_{\mathrm{opt}}$ & 1.0502 & 1.0254 & 1.0171 & 1.0129 & 1.0103 & 1.0086 & 1.0074 & 1.0065 & 1.0058 & 1.0052 \\
\bottomrule
\end{tabular}
\end{table}

Several trends are evident:
\begin{enumerate}
    \item $\beta_{\mathrm{opt}} > 1$ for all states and screening parameters, confirming that the Kratzer reference wave functions are slightly too diffuse and need to be contracted.
    \item The deviation of $\beta_{\mathrm{opt}}$ from unity generally decreases with $n$ for a fixed $c$, reflecting the reduced sensitivity of more diffuse excited states to short range screening.
    \item The ground state shows a large deviation from $\beta=1$, requiring the most significant adjustment.
\end{enumerate}

\subsection{Convergence Analysis: Error Scaling with Principal Quantum Number}
\begin{figure}[h]
\centering
\includegraphics[width=0.82\textwidth]{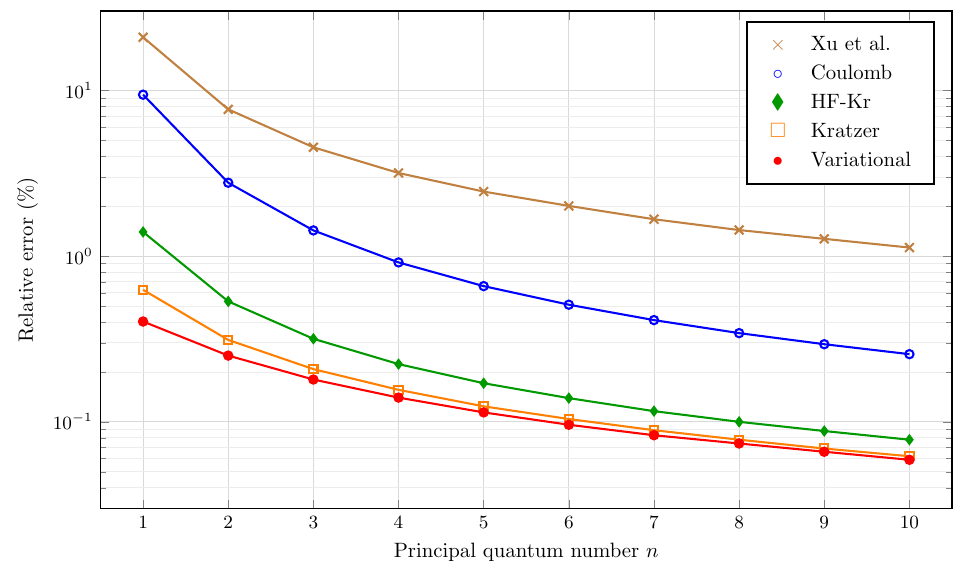}
\caption{Relative errors (\%, logarithmic scale) of the calculated energies for the first ten $s$ states of the hydrogen atom in the RSCP at $c=0.1$, measured with respect to the GPS reference values.}
\label{fig:fig1}
\end{figure}

Figure \ref{fig:fig1} presents the relative errors as functions of the principal quantum number $n$ for fixed $c = 0.1$, demonstrating the convergence behaviour of the different methods.
%=============================================================================
%=============================================================================
The error scaling reveals important convergence properties:
\begin{enumerate}
    \item \textbf{Coulomb and HF-C methods:} Errors decrease roughly as $1/n$ for large $n$, reflecting the diminishing importance of short range screening for more diffuse excited states. The ground state error is anomalously large due to the significant probability density at small $r$.
    \item \textbf{HF-Kr and EVA with Kratzer basis:} Errors are approximately constant or slowly varying with $n$, indicating that the effective angular momentum correction captures the dominant screening effect across all states.
    \item \textbf{Variational method:} Errors decrease smoothly with $n$ from 0.40\% for $1s$ to 0.06\% for $10s$, demonstrating the robustness of this approach across all states.
    \item \textbf{Xu et al. asymptotic approximation:}Errors decrease systematically with n, consistent with its asymptotic nature the approximation improves for higher excited states where the effective quantum number $N_\nu \gg 1$.
\end{enumerate}
The complementary error behaviour of our methods and the asymptotic approximation of Xu \textit{et al.}~\cite{Xu2024} suggests that different approaches may be optimal for different regimes: our Kratzer based methods for low lying states, and the asymptotic formula for highly excited states.

\begin{figure}[h]
\centering
\includegraphics[width=0.92\textwidth]{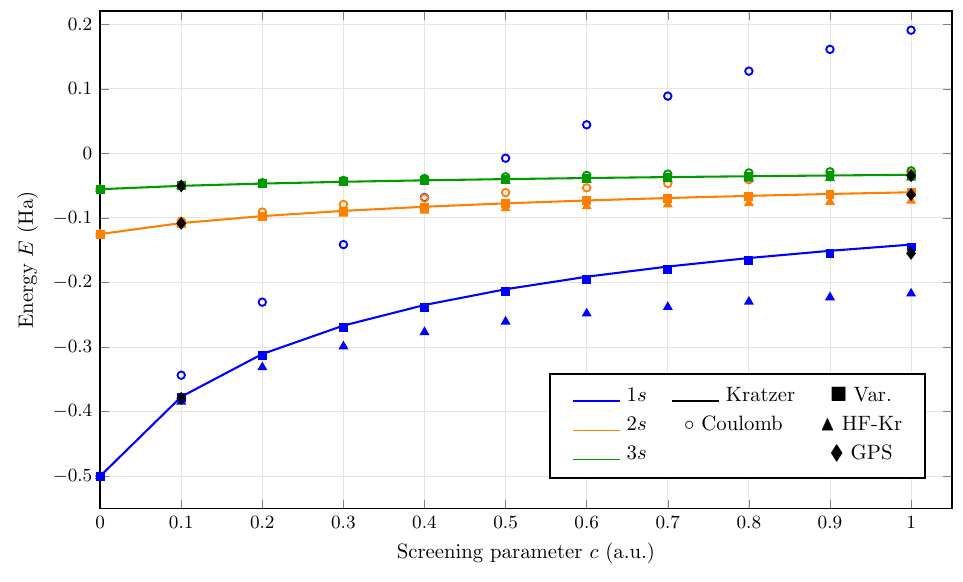}
\caption{Comprehensive comparison of energy eigenvalues for the first three $s$-states ($1s$, $2s$, $3s$) of the hydrogen atom in the RSCP as functions of the screening parameter $c$. All methods are shown for $c$ values from $0$ to $1$ with steps of $0.1$. GPS reference values are shown at both $c=0.1$ and $c=1.0$ (diamonds). Symbols: solid lines = Kratzer reference; open circles ($\circ$) = Coulomb reference; filled squares ($\blacksquare$) = variational method; filled triangles ($\blacktriangle$) = Hellmann--Feynman with Kratzer basis; filled diamonds ($\blacklozenge$) = GPS reference.}
\label{fig:fig2}
\end{figure}

\begin{figure}[h]
\centering
\includegraphics[width=0.92\textwidth]{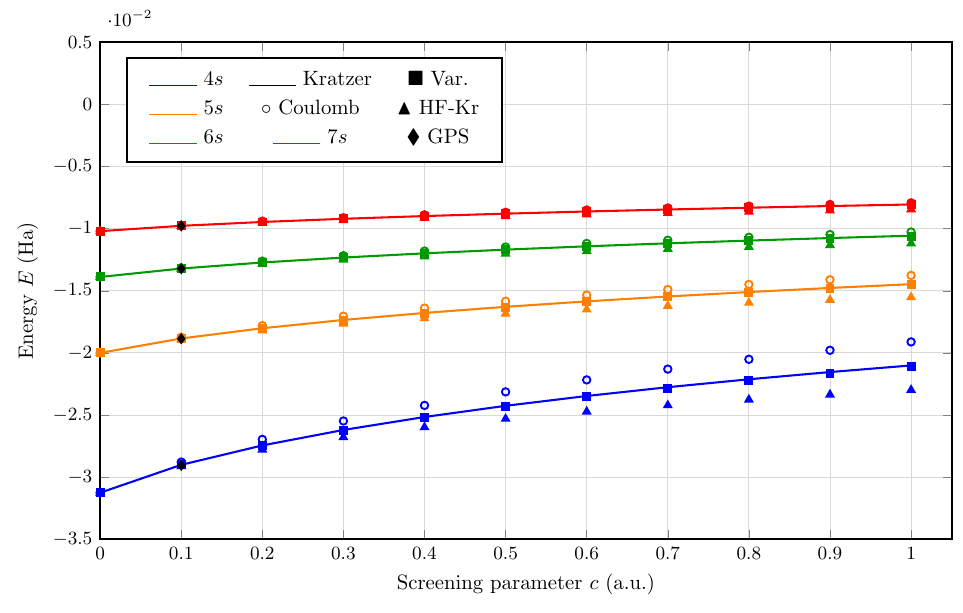}
\caption{Comprehensive comparison of energy eigenvalues for higher excited $s$ states ($4s$, $5s$, $6s$, and $7s$) of the hydrogen atom in the RSCP as functions of the screening parameter $c$. All methods are shown for $c$ values from $0$ to $1$ with steps of $0.1$. GPS reference values are shown at $c=0.1$ (diamonds). Symbols: solid lines = Kratzer reference; open circles ($\circ$) = Coulomb reference; filled squares ($\blacksquare$) = variational method; filled triangles ($\blacktriangle$) = Hellmann--Feynman with Kratzer basis; filled diamonds ($\blacklozenge$) = GPS reference.}
\label{fig:fig3}
\end{figure}

Figures \ref{fig:fig2} and \ref{fig:fig3} present a comprehensive comparison of all methods developed in this work for the first seven $s$-states of the hydrogen atom in the RSCP.
%=============================================================================
%=============================================================================
%=============================================================================
An interesting observation from Figure \ref{fig:fig2} concerns the apparent critical screening behaviour exhibited by the Coulomb reference method. The Coulomb based energy for $1s$ state crosses zero at approximately $c \approx 0.6$, suggesting a critical screening parameter for this bound state. However, this is an artifact of the approximation method, not a physical feature of the RSCP. The GPS calculations of Xu \textit{et al.}~\cite{Stachura2021,Xu2024} demonstrate that bound states persist to much larger values of $c$. The absence of critical screening in the exact RSCP can be understood by considering the effective potential including the centrifugal term:
\begin{equation}
V_{\text{eff}}(r) = \frac{l(l+1)}{2r^2} - \frac{e^{-c/r}}{r}.
\end{equation}
For $s$-states ($l=0$), the effective potential is purely attractive for all $r>0$, and as $r \to \infty$, it approaches the unscreened Coulomb form $-1/r$. This ensures that bound states exist for all finite values of $c$, although they become progressively less bound as $c$ increases.

The Kratzer based methods do not exhibit critical behaviour because they incorporate the $c/r^2$ screening correction into the reference Hamiltonian. This is consistent with the analysis of Xu \textit{et al.}~\cite{Xu2024},who showed that the effective potential approach (RSCP with modified centrifugal term) does not predict critical screening, in contrast to earlier work by Stachura~\cite{Stachura2021} who analysed the pure RSCP without the centrifugal contribution.
%=============================================================================
\section{Comparison with Xu et al. Asymptotic Approximation}
\label{sec:comparison_xu}
%=============================================================================
Xu \textit{et al.}~\cite{Xu2024} proposed the following asymptotic approximation for the eigenvalues of the RSCP:
\begin{equation}
E_{n,l}^{(\text{Xu})} = -\frac{1}{2(n+L-l)^2}, \quad L = -\frac{1}{2} + \sqrt{2c + \left(l+\frac{1}{2}\right)^2}.
\end{equation}
This approximation coincides with the bare Kratzer reference eigenvalue in Eq.~(\ref{eq:energy_final_Kr}), $E_{n_r,l}^{\text{Kr}} = -1/2N_\nu^2$ with $N_\nu = n_r + \nu_l + 1 = n + \nu_l - l$ and $\nu_l = -\frac{1}{2} + \sqrt{2c + (l+\frac{1}{2})^2}$. It does not include the additional expectation-value terms that arise when the full RSCP Hamiltonian is evaluated in the Kratzer reference state in Eq.~(\ref{eq:energy_final_Kr}).

\begin{figure}[h]
\centering
\includegraphics[width=0.92\textwidth]{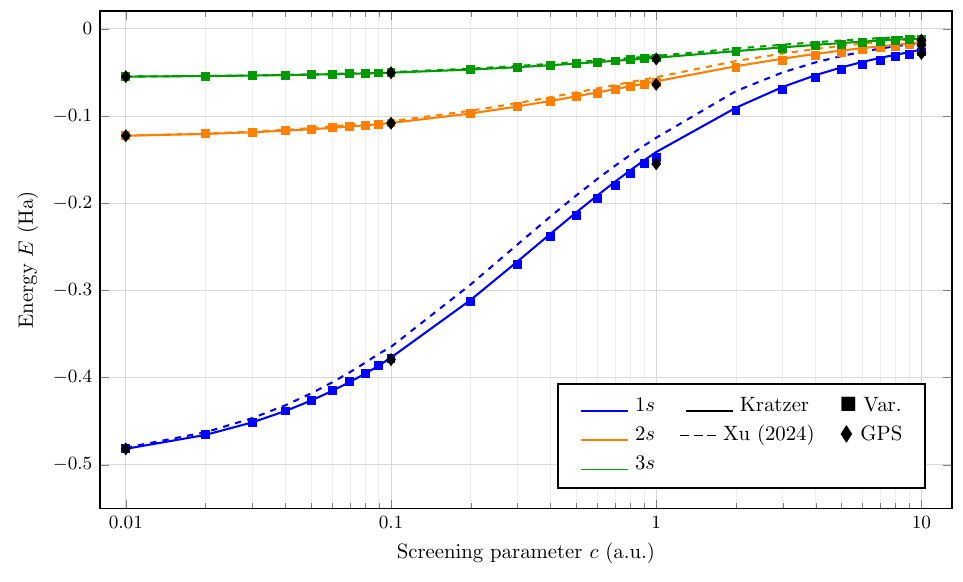}
\caption{Comparison of the EVA with Kratzer basis (solid lines) with the asymptotic approximation of Xu \textit{et al.} (2024) (dashed lines) for the first three $s$-states over the range $0 \leq c \leq 1$. Filled squares ($\blacksquare$) show the variational results. Diamonds ($\blacklozenge$) indicate GPS reference values at $c=0.1$ and $c=1.0$.}
\label{fig:fig4}
\end{figure}

\begin{figure}[h]
\centering
\includegraphics[width=0.92\textwidth]{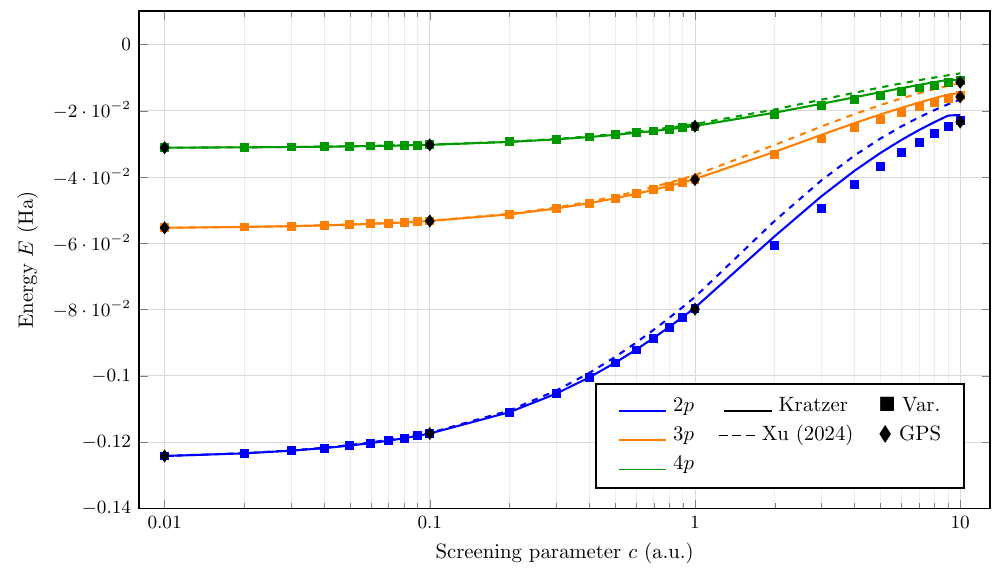}
\caption{Comparison of the EVA with Kratzer basis (solid lines) with the asymptotic approximation of Xu \textit{et al.} (2024) (dashed lines) for the $p$-states ($2p$, $3p$, $4p$) over the range $0 \leq c \leq 1$. Filled squares ($\blacksquare$) show the variational results. Diamonds ($\blacklozenge$) indicate GPS reference values at $c=0.1$ and $c=1.0$.}
\label{fig:fig5}
\end{figure}

Figure \ref{fig:fig4} presents a comparison of our expectation value approach with Kratzer basis, the variational method, and the Xu \textit{et al.} asymptotic approximation over the screening range $0 \le c \le 1$ for $s$-states.
Figure \ref{fig:fig5} presents the same comparison for the $p$-states ($2p, 3p, 4p$) that are available in the GPS data from Xu \textit{et al.}~\cite{Xu2024}.
%=============================================================================
%=============================================================================
%=============================================================================
The comparison demonstrates that:
\begin{itemize}
    \item Our expectation value approach with Kratzer basis and the Xu \textit{et al.}asymptotic approximation are complementary approaches with different strengths.
    \item For low ying states (especially $1s$), our method provides improved accuracy by including the expectation value corrections.
    \item For highly excited states, both methods converge to similar accuracy, consistent with the asymptotic nature of the Xu \textit{et al.} formula.
    \item The variational method provides the best accuracy across all states and screening parameters.
    \item At large $c$ (e.g., $c=1.0$), both analytical methods show increased deviation from GPS values, indicating the limitations of first order approximations in the strong screening regime  higher order corrections in the Taylor expansion of RSCP or alternative resummation techniques may be needed for accurate
results in this regime.
    \item For $p$-states (Figure \ref{fig:fig5}), the differences between methods are smaller than for $s$-states because the centrifugal barrier reduces the wave function amplitude at small $r$ where screening effects are strongest.
\end{itemize}
The complementary nature of the two approaches suggests that practitioners may choose between them depending on the specific application: our expectation value approach with Kratzer basis for accurate low lying state energies, and the Xu \textit{et al.} asymptotic formula for quick estimates of highly excited state energies or for analytical studies where simplicity is paramount.
%=============================================================================
\section{Extension to Positronium}
\label{sec:positronium}
%=============================================================================
The formalism developed in Sections 3--5 for hydrogen can be extended to any hydrogen-like system with an arbitrary reduced mass. In this section, we apply our methods to Positronium (Ps), the bound state of an electron and a positron.

For Positronium, the reduced mass is $\mu_{\text{Ps}} = m_e/2 = 1/2$ in atomic units. This leads to the following modifications compared to hydrogen:
\begin{itemize}
    \item The effective Bohr radius is $a_{\text{Ps}} = 2a_0 = 2$ a.u. (twice that of hydrogen).
    \item The unscreened ground state energy is $E_{1s}^{(0),\text{Ps}} = -1/4$ Ha (half that of hydrogen).
    \item The effective angular momentum becomes:
    \begin{equation}
    \nu_l^{\text{Ps}} = -\frac{1}{2} + \sqrt{\left(l+\frac{1}{2}\right)^2+c},
    \end{equation}
    satisfying $\nu^{\text{Ps}}(\nu^{\text{Ps}}+1) = l(l+1)+c$.
\end{itemize}
Table \ref{tab:ps_energies_errors_c01} presents the energy eigenvalues for the first ten $s$-states of Positronium in the RSCP at $c=0.1$, computed using our analytical methods and compared to reference values.
%=============================================================================
\begin{table}[h]
\centering
\caption{Energy eigenvalues (in Ha) for the first ten $s$ states of Positronium in the RSCP at $c=0.1$. Relative errors with respect to the reference values of~\cite{Xu2024}, expressed in percent, are given in parentheses.}
\label{tab:ps_energies_errors_c01}
\resizebox{\textwidth}{!}{%
\begin{tabular}{l c c c c c}
\toprule
\textbf{State} & \textbf{Coulomb ref.} & \textbf{HF-Kr} & \textbf{Kratzer ref.} & \textbf{Variational} & \textbf{S-H (2021)~\cite{Stachura2021}} \\
\midrule
1s  & $-2.069571(-1)$ (3.005) & $-2.141333(-1)$ (0.358) & $-2.130439(-1)$ (0.153) & $-2.131429(-1)$ (0.106) & $-2.134000(-1)$ (0.014) \\
2s  & $-5.710253(-2)$ (0.935) & $-5.772139(-2)$ (0.139) & $-5.759867(-2)$ (0.074) & $-5.760561(-2)$ (0.062) & $-5.760000(-2)$ (0.072) \\
3s  & $-2.617752(-2)$ (0.499) & $-2.633080(-2)$ (0.083) & $-2.629593(-2)$ (0.049) & $-2.629736(-2)$ (0.044) & $-2.630000(-2)$ (0.034) \\
4s  & $-1.494974(-2)$ (0.328) & $-1.500784(-2)$ (0.059) & $-1.499346(-2)$ (0.037) & $-1.499392(-2)$ (0.034) & $-1.500000(-2)$ (0.007) \\
5s  & $-9.654232(-3)$ (0.240) & $-9.681924(-3)$ (0.046) & $-9.674665(-3)$ (0.029) & $-9.674858(-3)$ (0.027) & $-9.700000(-3)$ (0.232) \\
6s  & $-6.744336(-3)$ (0.188) & $-6.759560(-3)$ (0.037) & $-6.755400(-3)$ (0.024) & $-6.755494(-3)$ (0.023) & $-6.800000(-3)$ (0.636) \\
7s  & $-4.976021(-3)$ (0.154) & $-4.985245(-3)$ (0.031) & $-4.982644(-3)$ (0.021) & $-4.982695(-3)$ (0.020) & $-5.000000(-3)$ (0.327) \\
8s  & $-3.821824(-3)$ (0.130) & $-3.827821(-3)$ (0.027) & $-3.826087(-3)$ (0.018) & $-3.826118(-3)$ (0.017) & $-3.800000(-3)$ (0.700) \\
9s  & $-3.027124(-3)$ (0.112) & $-3.031235(-3)$ (0.024) & $-3.030023(-3)$ (0.016) & $-3.030042(-3)$ (0.016) & $-3.000000(-3)$ (1.007) \\
10s & $-2.456773(-3)$ (0.098) & $-2.459711(-3)$ (0.021) & $-2.458830(-3)$ (0.015) & $-2.458843(-3)$ (0.014) & $-2.500000(-3)$ (1.660) \\
\bottomrule
\end{tabular}
}
\end{table}
%=============================================================================
Table \ref{tab:H_Ps_ground} compares the ground-state energies of hydrogen and Positronium at $c=0.1$. The fractional energy shift for hydrogen ($24.44\%$) is larger than for positronium ($14.74\%$) at the same value of $c$. This difference arises from the modified effective angular momentum: $\nu_0^{\text{H}} \approx 0.171$ versus $\nu_0^{\text{Ps}} \approx 0.092$ at $c=0.1$. Physically, hydrogen experiences a larger effective centrifugal barrier modification for the same screening parameter, leading to stronger binding reduction.
%=============================================================================
\begin{table}[h]
\centering
\caption{Comparison of ground state energies (in Ha) for hydrogen and Positronium in the RSCP at $c = 0.1$ with variational method. The last column gives the relative energy shift with respect to the unscreened ground-state energy.}
\label{tab:H_Ps_ground}
\begin{tabular}{l c c c}
\toprule
\textbf{System} & $\boldsymbol{E_{1s}^{(0)}}$ & $\boldsymbol{E_{1s}(c=0.1)}$ & $\boldsymbol{\Delta E / E^{(0)}}$ (\%) \\
\midrule
Hydrogen    & $-5.00(-1)$ & $-3.778187(-1)$ & $24.44$ \\
Positronium & $-2.50(-1)$ & $-2.131429(-1)$ & $14.74$ \\
\bottomrule
\end{tabular}
\end{table}
%=============================================================================
\section{Conclusions}
\label{sec:conclusions}
%=============================================================================
In this paper, we have developed a comprehensive analytical framework for approximating the eigenenergies of hydrogen-like atoms confined in a plasma medium characterised by the radial screened Coulomb potential (RSCP). Three complementary approaches - expectation-value with Coulomb and Kratzer basis, variational, and Hellmann-Feynman theorem - yield closed-form expressions in terms of modified Bessel functions, valid for arbitrary quantum numbers $n$ and $\ell$. The Kratzer-based method significally improve accuracy by incorporating the leading $c/r^{2}$ correction via effective angular momentum, with the variational method achieving the highest precision (errors $<0.41\%$ for the ten $s$-states at $c=0.1$)

The extension to positronium reveals mass-dependent screening-responses, with hydrogen experiencing a larger binding redution than positronium for the same screening parameter due to its larger effective centrifugal barrier modification. While all analytical methods show inceased deviations for strong screening ($c\geq 1$), the present toolkit provides robust, efficeint, and accurate estimates across a wide range of screening strengths. Future refinements may include multi-parameter optimization, hugh-order resummation techniques, and extensions to more complex atomic systems.

%=============================================================================
\section*{Appendices}
%=============================================================================

\appendix

\section{Unified Derivation of Expectation Values with Generalised Kratzer Basis}
\label{app:A}

We present a unified derivation of the RSCP energy expectation value using a two parameter family of trial wave functions.  An artificial parameter $k\in[0,1]$ interpolates between the Coulomb ($k=0$) and Kratzer ($k=1$)
reference potentials, while a scaling parameter $\beta>0$ controls the spatial extent of the wave function.  All three methods developed in the main text Coulomb reference, Kratzer reference, and variational are recovered as special cases of a single master formula.

Consider the reference potential
\begin{equation}\label{eq:Vref-k}\tag{A.1}
  V_{\mathrm{ref}}^{(k)}(r) = -\frac{1}{r} + \frac{k\,c}{r^{2}}\,,
\end{equation}
The corresponding reference Hamiltonian is
\begin{equation}\label{eq:Href-k}\tag{A.2}
  H_{\mathrm{ref}}^{(k)}
  = -\frac{1}{2}\frac{d^{2}}{dr^{2}}
    + \frac{\nu_{k}(\nu_{k}+1)}{2r^{2}}
    - \frac{1}{r}\,,
\end{equation}
where the effective angular momentum quantum number is defined by
\begin{equation}\label{eq:nuk}\tag{A.3}
  \nu_{k} = -\frac{1}{2}
    + \sqrt{\Bigl(l+\tfrac{1}{2}\Bigr)^{2} + 2kc}\,,
\end{equation}
satisfying $\nu_{k}(\nu_{k}+1) = l(l+1)+2kc$.  The key limiting cases are:
\begin{itemize}
  \item $k=0$:\; $\nu_{0}=l$\quad (Coulomb);
  \item $k=1$:\; $\nu_{1}=-\tfrac{1}{2}+\sqrt{(l+\tfrac{1}{2})^{2}+2c}
        \equiv\nu$\quad (Kratzer).
\end{itemize}

The eigenvalues and normalised reduced radial eigenfunctions of
$H_{\mathrm{ref}}^{(k)}$ are
\begin{equation}\label{eq:Ek}\tag{A.4}
  E_{n_{r}}^{(k)} = -\frac{1}{2N_{k}^{2}}\,,
  \qquad
  N_{k} = n_{r}+\nu_{k}+1\,,
\end{equation}
\begin{equation}\label{eq:Pk}\tag{A.5}
  P_{n_{r},l}^{(k)}(r)
  = \frac{1}{N_{k}}
    \sqrt{\frac{\Gamma(n_{r}+1)}{\Gamma(n_{r}+2\nu_{k}+2)}}
    \left(\frac{2r}{N_{k}}\right)^{\!\nu_{k}+1}
    e^{-r/N_{k}}\,
    L_{n_{r}}^{2\nu_{k}+1}\!\!\left(\frac{2r}{N_{k}}\right),
\end{equation}
where for non-integer $\nu_{k}$ the Laguerre polynomials are defined through the generalised series representation and factorials are replaced by Gamma functions.  The standard expectation values are
\begin{equation}\label{eq:ev-k}\tag{A.6}
  \left\langle\frac{1}{r}\right\rangle_{\!k}
  = \frac{1}{N_{k}^{2}}\,,
  \qquad
  \left\langle\frac{1}{r^{2}}\right\rangle_{\!k}
  = \frac{1}{N_{k}^{3}\bigl(\nu_{k}+\tfrac{1}{2}\bigr)}\,.
\end{equation}

We construct the two parameter trial wave function by introducing a scaling parameter $\beta>0$:
\begin{equation}\label{eq:Ptrial}\tag{A.7}
  P_{n_{r},l}^{\mathrm{trial}}(r;\,k,\beta)
  = \mathcal{N}(k,\beta)\,
    \left(\frac{2\beta r}{N_{k}}\right)^{\!\nu_{k}+1}
    e^{-\beta r/N_{k}}\,
    L_{n_{r}}^{2\nu_{k}+1}\!\!\left(\frac{2\beta r}{N_{k}}\right).
\end{equation}
$\beta=1$ recovers the eigenfunction~(\ref{eq:Pk});
$\beta>1$ contracts the wave function and $\beta<1$ expands it.

Introducing $\rho=2\beta r/N_{k}$, the normalisation integral becomes
\begin{equation}\label{eq:norm-int}\tag{A.8}
  \int_{0}^{\infty}\!|P_{n_{r},l}^{\mathrm{trial}}|^{2}\,dr
  = |\mathcal{N}|^{2}\,\frac{N_{k}}{2\beta}
    \int_{0}^{\infty}\!\rho^{2\nu_{k}+2}\,e^{-\rho}\,
    \bigl[L_{n_{r}}^{2\nu_{k}+1}(\rho)\bigr]^{2}\,d\rho
  = 1\,.
\end{equation}
Using the standard Laguerre identity
\begin{equation}\label{eq:Lag-orth}\tag{A.9}
  \int_{0}^{\infty}\!\rho^{\alpha+1}\,e^{-\rho}\,
  \bigl[L_{n}^{\alpha}(\rho)\bigr]^{2}\,d\rho
  = (2n+\alpha+1)\,\frac{\Gamma(n+\alpha+1)}{\Gamma(n+1)}\,,
\end{equation}
with $\alpha=2\nu_{k}+1$ and $n=n_{r}$, we obtain
\begin{equation}\label{eq:Nkb}\tag{A.10}
  \mathcal{N}(k,\beta)
  = \frac{\sqrt{\beta}}{N_{k}}
    \sqrt{\frac{\Gamma(n_{r}+1)}{\Gamma(n_{r}+2\nu_{k}+2)}}\,.
\end{equation}

The full RSCP Hamiltonian is
\begin{equation}\label{eq:HRSCP}\tag{A.11}
  H = -\frac{1}{2}\frac{d^{2}}{dr^{2}}
    + \frac{l(l+1)}{2r^{2}}
    - \frac{e^{-c/r}}{r}\,.
\end{equation}
The energy expectation value with the trial function~(\ref{eq:Ptrial}) decomposes as
\begin{equation}\label{eq:Edecomp}\tag{A.12}
  E(k,\beta)
  = \bigl\langle\hat{T}+\hat{V}_{\mathrm{eff}}\bigr\rangle_{k,\beta}
    + \langle V_{\mathrm{RSCP}}\rangle_{k,\beta}\,.
\end{equation}

For the scaled wave function, expectation values scale as
$\langle r^{-j}\rangle_{k,\beta}=\beta^{j}\langle r^{-j}\rangle_{k,1}$
and
$\langle d^{2}/dr^{2}\rangle_{k,\beta}=\beta^{2}\langle d^{2}/dr^{2}\rangle_{k,1}$.
Using the eigenvalue relation for the unscaled ($\beta=1$) Kratzer Hamiltonian,
\begin{equation}\label{eq:virial-k}\tag{A.13}
  \langle\hat{T}\rangle_{k,1}
  + \frac{\nu_{k}(\nu_{k}+1)}{2}
    \left\langle\frac{1}{r^{2}}\right\rangle_{\!k,1}
  = \frac{1}{2N_{k}^{2}}\,,
\end{equation}
the physical kinetic plus centrifugal expectation value becomes ($\nu_{k}(\nu_{k}+1)=l(l+1)+2kc$)
\begin{align}
  \bigl\langle\hat{T}+\hat{V}_{\mathrm{eff}}\bigr\rangle_{k,\beta}
  &= \beta^{2}\!\left(
       \langle\hat{T}\rangle_{k,1}
       + \frac{l(l+1)}{2}\left\langle\frac{1}{r^{2}}\right\rangle_{\!k,1}
     \right)
  = \beta^{2}\!\left(
       \frac{1}{2N_{k}^{2}}
       - kc\left\langle\frac{1}{r^{2}}\right\rangle_{\!k,1}
     \right) \nonumber \notag \\[4pt]
  &= \frac{\beta^{2}}{2N_{k}^{2}}
     \left(1-\frac{2kc}{N_{k}\bigl(\nu_{k}+\tfrac{1}{2}\bigr)}\right). \label{eq:TplusVeff}\tag{A.14}
\end{align}
Using the substitution $u=\beta r$ and the scaling property
$|P^{\mathrm{trial}}(r;\,k,\beta)|^{2}\,dr=|P^{(k)}(u)|^{2}\,du$, we obtain
\begin{equation}\label{eq:VRSCP-scale}\tag{A.15}
  \langle V_{\mathrm{RSCP}}\rangle_{k,\beta}
  = -\beta\!\int_{0}^{\infty}\!|P_{n_{r}}^{(k)}(u)|^{2}\,
    \frac{e^{-c\beta/u}}{u}\,du
  = -\beta\left\langle\frac{e^{-c\beta/r}}{r}\right\rangle_{\!k,1},
\end{equation}
which is $\beta$ times the unscaled expectation value evaluated with effective screening parameter $c'=c\beta$.

To evaluate $\langle e^{-c'/r}/r\rangle_{k,1}$, we introduce $\rho=2r/N_{k}$ and expand the generalised Laguerre polynomial:
\begin{equation}\label{eq:Lag-expand}\tag{A.16}
  L_{n_{r}}^{2\nu_{k}+1}(\rho)
  = \sum_{m=0}^{n_{r}}
    \frac{(-1)^{m}\,\Gamma(n_{r}+2\nu_{k}+2)}
         {\Gamma(m+1)\,\Gamma(n_{r}-m+1)\,\Gamma(2\nu_{k}+2+m)}\,
    \rho^{m}\,.
\end{equation}
After squaring and applying the master integral formula
\begin{equation}\label{eq:master}\tag{A.17}
  \int_{0}^{\infty}\!x^{\mu-1}\,e^{-\gamma x-\delta/x}\,dx
  = 2\!\left(\frac{\delta}{\gamma}\right)^{\!\mu/2}
    K_{\mu}\!\bigl(2\sqrt{\delta\gamma}\bigr)\,,
  \qquad
  \Re(\delta)>0,\;\Re(\gamma)>0\,,
\end{equation}
with $\gamma=1$ and $\delta=2c'/N_{k}$, we obtain the general screened expectation value:
\begin{align}\label{eq:ecr-general}
  \left\langle\frac{e^{-c'/r}}{r}\right\rangle_{\!k,1}
  &= \frac{2\,\Gamma(n_{r}+1)\,\Gamma(n_{r}+2\nu_{k}+2)}{N_{k}^{2}}
     \sum_{m=0}^{n_{r}}\sum_{m'=0}^{n_{r}}
     \chi_{mm'}^{(n_{r},\nu_{k})} \nonumber \notag\\[2pt]
  &\quad\times
     \left(\frac{2c'}{N_{k}}\right)^{\!(2\nu_{k}+2+m+m')/2}
     K_{2\nu_{k}+2+m+m'}\!\!
     \left(2\sqrt{\frac{2c'}{N_{k}}}\right), \tag{A.18}
\end{align}
where the coefficients are
\begin{equation}\label{eq:chi-coeff} \tag{A.19}
  \chi_{mm'}^{(n_{r},\nu_{k})}
  = \frac{(-1)^{m+m'}}
         {\Gamma(m\!+\!1)\,\Gamma(m'\!+\!1)\,
          \Gamma(n_{r}\!-\!m\!+\!1)\,\Gamma(n_{r}\!-\!m'\!+\!1)\,
          \Gamma(2\nu_{k}\!+\!2\!+\!m)\,\Gamma(2\nu_{k}\!+\!2\!+\!m')}\,.
\end{equation}
Combining Eqs.~(\ref{eq:TplusVeff}) and~(\ref{eq:VRSCP-scale}) with~(\ref{eq:ecr-general}), the complete RSCP energy with parameters $(k,\beta)$ is
\begin{equation}\label{eq:Ekbeta} \tag{A.20}
    E_{n_{r},l}(k,\beta)
  = \frac{\beta^{2}}{2N_{k}^{2}}
    \!\left(1-\frac{2kc}{N_{k}\bigl(\nu_{k}+\tfrac{1}{2}\bigr)}\right)
  - \frac{\beta\,\Gamma(n_{r}+1)}
         {N_{k}^{2}\,\Gamma(n_{r}+2\nu_{k}+2)}\;
    \mathcal{I}_{n_{r},l}^{\mathrm{RSCP}}(c\beta;\,k)
\end{equation}
where
\begin{align}\label{eq:IRSCP-def} \tag{A.21}
  \mathcal{I}_{n_{r},l}^{\mathrm{RSCP}}(c';\,k)
  = 2\,\bigl[\Gamma(n_{r}+2\nu_{k}+2)\bigr]^{2}
     \sum_{m=0}^{n_{r}}\sum_{m'=0}^{n_{r}}
     \chi_{mm'}^{(n_{r},\nu_{k})}
     \left(\frac{2c'}{N_{k}}\right)^{\nu_{k}+1+\frac{m+m'}{2}}
     K_{2\nu_{k}+2+m+m'}\!\!
     \left(2\sqrt{\frac{2c'}{N_{k}}}\right).
\end{align}

The unified expression~(\ref{eq:Ekbeta}) generates all methods developed in the main text through specific choices of $(k,\beta)$.

\subsection{Case~1: Coulomb reference  ($k=0$, $\beta=1$)}

Setting $k=0$ gives $\nu_{k}=l$, $N_{k}=n_{r}+l+1=n$, and the first term in Eq.~(\ref{eq:Ekbeta}) simplifies to $1/(2n^{2})$.
The Laguerre polynomials reduce to the standard associated Laguerre polynomials $L_{n-l-1}^{2l+1}$ with integer indices, and the coefficients $\chi_{mm'}^{(n_{r},l)}$ involve ordinary factorials. The energy recovers Eq.~(\ref{eq:energy_coulomb_decomp}) of the main text and becomes
\begin{equation}\label{eq:EC-recover} \tag{A.22}
  E_{n,l}^{\mathrm{C}}
  = \frac{1}{2n^{2}}
    - \left\langle\frac{e^{-c/r}}{r}\right\rangle_{\!n,l}^{\!\mathrm{C}}\,,
\end{equation}

\subsection{Case~2: Kratzer reference ($k=1$, $\beta=1$)}

Setting $k=1$ gives
$\nu_{k}=\nu=-\tfrac{1}{2}+\sqrt{(l+\tfrac{1}{2})^{2}+2c}$ and
$N_{k}=N_{\nu}=n_{r}+\nu+1$, so the energy becomes
\begin{equation}\label{eq:EKr-recover}\tag{A.23}
  E_{n_{r},l}^{\mathrm{Kr}}
  = \frac{1}{2N_{\nu}^{2}}
    \!\left(1-\frac{2c}{N_{\nu}\bigl(\nu+\tfrac{1}{2}\bigr)}\right)
  - \left\langle\frac{e^{-c/r}}{r}\right\rangle_{\!n_{r},l}^{\!\mathrm{Kr}}.
\end{equation}
Using Eq.~(\ref{eq:ev-k}), the first term can be rewritten as
\begin{equation} \tag{A.24}
  \frac{1}{2N_{\nu}^{2}}
  - \frac{c}{N_{\nu}^{3}\sqrt{(l+\tfrac{1}{2})^{2}+2c}}\,,
\end{equation}
recovering Eq.~(\ref{eq:energy_final_Kr}) of the main text.  The Bessel function indices $2\nu_{k}+2+m+m'$ become $2\nu+2+m+m'$.

\subsection{Case~3: Variational method( $k=1$, $\beta=\beta_{\mathrm{opt}}$)}

With $k=1$, the energy $E_{n_{r},l}(1,\beta)$ from Eq.~(\ref{eq:Ekbeta}) is a function of the single variational parameter $\beta$.  The optimal $\beta_{\mathrm{opt}}$ is determined by
\begin{equation}\label{eq:beta-opt}\tag{A.25}
  \frac{\partial E_{n_{r},l}(1,\beta)}{\partial\beta}
  \bigg|_{\beta=\beta_{\mathrm{opt}}} = 0\,.
\end{equation}
Due to the transcendental dependence on $\beta$ through the modified Bessel functions (with argument $\tilde{z}=2\sqrt{2c\beta/N_{\nu}}$), this equation is solved numerically.  The variational energy
$E_{n_{r},l}(1,\beta_{\mathrm{opt}})$ provides the best estimate within the scaled Kratzer ansatz, as discussed in Section~\ref{sec:variational}.

For the ground state ($n_{r}=0$, $l=0$), only the $m=m'=0$ term survives:
\begin{equation}\label{eq:E1s-var} \tag{A.26}
  E_{\mathrm{var}}^{(1s)}(\beta)
  = \frac{\beta^{2}}{2N_{\nu}^{2}}
    \!\left(1-\frac{2c}{N_{\nu}\bigl(\nu+\tfrac{1}{2}\bigr)}\right)
  - \frac{2\beta\,\tilde{c}^{\,\nu+1}}
         {N_{\nu}^{2}\,\Gamma(2\nu+2)}\,
    K_{2\nu+2}(\tilde{z})\,,
\end{equation}
where $\tilde{c}=2c\beta/N_{\nu}$ and $\tilde{z}=2\sqrt{\tilde{c}}$, in agreement with Eq.~(\ref{eq:total_var_energy}) of the main text.

In the limit $c\to 0$ (for any~$k$), we have $\nu_{k}\to l$, $N_{k}\to n$, and the small argument expansion
$K_{\mu}(z)\sim\tfrac{\Gamma(\mu)}{2}(2/z)^{\mu}$ ensures
\begin{equation}\label{eq:c-to-0}\tag{A.27}
  \left\langle\frac{e^{-c/r}}{r}\right\rangle_{\!k,1}
  \;\longrightarrow\;
  \left\langle\frac{1}{r}\right\rangle = \frac{1}{n^{2}}\,,
\end{equation}
so that the energy reduces to the unscreened Coulomb value $E_{n}^{(0)}=-1/(2n^{2})$ for all three methods.

Similarly, in the limit $c\to 0$ with $\beta$ fixed:
\begin{equation}\label{eq:virial-check}\tag{A.28}
  E(k,\beta)
  \;\longrightarrow\;
  \frac{\beta^{2}}{2n^{2}} - \frac{\beta}{n^{2}}
  = -\frac{1}{2n^{2}} + \frac{(\beta-1)^{2}}{2n^{2}}\,,
\end{equation}
confirming that $\beta_{\mathrm{opt}}\to 1$ as $c\to 0$.

\subsection{Summary: parameter choices and corresponding methods}
\label{app:summary}

Table~\ref{tab:summary-params} collects the parameter assignments for all three methods and their correspondence to the equations in the main text.

\begin{table}[h]
\centering
\caption{Parameter choices $(k,\beta)$ in the unified energy
expression~(\ref{eq:Ekbeta}) and the corresponding methods.}
\label{tab:summary-params}
\begin{tabular}{lccll}
\hline
Method & $k$ & $\beta$ & Energy expression & Main text Eq. \\
\hline
Coulomb reference (EVA)
  & $0$ & $1$
  & $E_{n,l}^{\mathrm{C}}
     = \dfrac{1}{2n^{2}}
       - \left\langle\dfrac{e^{-c/r}}{r}\right\rangle_{\!n,l}^{\!\mathrm{C}}$
  & (\ref{eq:energy_coulomb_decomp})\\[12pt]
Kratzer reference (EVA)
  & $1$ & $1$
  & $E_{n_{r},l}^{\mathrm{Kr}}
     = \dfrac{1}{2N_{\nu}^{2}}
       \!\left(1\!-\!\dfrac{2c}{N_{\nu}(\nu+\frac{1}{2})}\right)
     - \left\langle\dfrac{e^{-c/r}}{r}\right\rangle_{\!n_{r},l}^{\!\mathrm{Kr}}$
  & (\ref{eq:energy_final_Kr}) \\[12pt]
Variational
  & $1$ & $\beta_{\mathrm{opt}}$
  & $E_{n_{r},l}^{\mathrm{var}}
     = E_{n_{r},l}(1,\beta_{\mathrm{opt}})$
     via $\partial E/\partial\beta=0$
  & (\ref{eq:total_var_energy})--(\ref{eq:optimization_condition}) \\
\hline
\end{tabular}
\end{table}

The unified framework makes transparent that all three methods share the same mathematical structure evaluation of the RSCP Hamiltonian expectation value with generalised Laguerre type wave functions and differ only in the degree of optimisation of the trial function parameters.

%-----------------------------------------------------------------------
\section{Hellmann-Feynman Energy Formula with Kratzer Basis}
\label{app:B}

We derive the Hellmann--Feynman expression for the energy derivative using the Kratzer reference eigenfunctions for arbitrary quantum numbers $(n_r,l)$. The theorem gives
\begin{equation}
\frac{\partial E_{n_r,l}}{\partial c} = \left\langle \frac{e^{-c/r}}{r^2} \right\rangle^{\mathrm{Kr}}_{n_r,l} = \int_0^\infty \left|P^{\mathrm{Kr}}_{n_r,l}(r) \right|^2 \frac{e^{-c/r}}{r^2}\,dr .
\label{eq:B1} \tag{B.1}
\end{equation}

The Kratzer reduced radial eigenfunction is ($\nu=-1/2+\sqrt{\left(l+1/2\right)^2+2c}$, $N_\nu=n_r+\nu+1$)
\begin{equation}
P^{\mathrm{Kr}}_{n_r,l}(r) = \frac{1}{N_\nu} \sqrt{\frac{\Gamma(n_r+1)}{\Gamma(n_r+2\nu+2)}} \left(\frac{2r}{N_\nu}\right)^{\nu+1} e^{-r/ N_\nu} L_{n_r}^{\,2\nu+1}\!\left(\frac{2r}{N_\nu}\right),
\label{eq:B2} \tag{B.2}
\end{equation}

Introducing the dimensionless variable $\rho=2r/N_\nu$, and using
\begin{equation}
\left|P^{\mathrm{Kr}}_{n_r,l}(r)\right|^2 dr = \frac{\Gamma(n_r+1)}{2N_\nu\,\Gamma(n_r+2\nu+2)} \,\rho^{2\nu+2}e^{-\rho} \left[L_{n_r}^{\,2\nu+1}(\rho)\right]^2 d\rho,
\label{eq:B3} \tag{B.3}
\end{equation}
we note that
\begin{equation}
\frac{e^{-c/r}}{r^2} = \frac{4}{N_\nu^2\,\rho^2} e^{-2c/(N_\nu\rho)}.
\label{eq:B4} \tag{B.4}
\end{equation}
Hence Eq.~\ref{eq:B1} becomes
\begin{equation}
\left\langle \frac{e^{-c/r}}{r^2} \right\rangle^{\mathrm{Kr}}_{n_r,l} = \frac{2\,\Gamma(n_r+1)}{N_\nu^3\,\Gamma(n_r+2\nu+2)} \,I^{\mathrm{HF}}_{n_r,l},
\label{eq:B5} \tag{B.5}
\end{equation}
with
\begin{equation}
I^{\mathrm{HF}}_{n_r,l} = \int_0^\infty \rho^{2\nu} e^{-\rho-2c/(N_\nu\rho)} \left[L_{n_r}^{\,2\nu+1}(\rho)\right]^2 d\rho .
\label{eq:B6} \tag{B.6}
\end{equation}

Using the same generalized Laguerre expansion as in Appendix~\ref{app:A},
\begin{equation}
L_{n_r}^{\,2\nu+1}(\rho) = \sum_{m=0}^{n_r} \frac{(-1)^m\,\Gamma(n_r+2\nu+2)} {\Gamma(m+1)\,\Gamma(n_r-m+1)\,\Gamma(2\nu+2+m)} \rho^m ,
\label{eq:B7} \tag{B.7}
\end{equation}
we obtain
\begin{equation}
\left[L_{n_r}^{\,2\nu+1}(\rho)\right]^2 = \Gamma(n_r+2\nu+2)^2 \sum_{m=0}^{n_r}\sum_{m'=0}^{n_r} \chi^{(n_r,\nu)}_{mm'}\,\rho^{m+m'}, 
\label{eq:B8} \tag{B.8}
\end{equation}
where the coefficients $\chi^{(n_r,\nu)}_{mm'}$ are
\begin{equation}
\chi^{(n_r,\nu)}_{mm'} = \frac{(-1)^{m+m'}}{\Gamma(m+1)\Gamma(m'+1)\Gamma(n_r-m+1)\Gamma(n_r-m'+1) \Gamma(2\nu+2+m) \Gamma(2\nu+2+m')}.
\label{eq:B9} \tag{B.9}
\end{equation}

Substituting Eq.~\ref{eq:B8} into Eq.~\ref{eq:B6}, we find
\begin{equation}
I^{\mathrm{HF}}_{n_r,l} = \Gamma(n_r+2\nu+2)^2 \sum_{m=0}^{n_r}\sum_{m'=0}^{n_r} \chi^{(n_r,\nu)}_{mm'}\int_0^\infty \rho^{2\nu+m+m'} e^{-\rho-2c/(N_\nu\rho)}\,d\rho, 
\label{eq:B10} \tag{B.10}
\end{equation}
Using the master integral
\begin{equation}
\int_0^\infty x^{\mu-1}e^{-\beta/x-\gamma x}\,dx = 2\left(\frac{\beta}{\gamma}\right)^{\mu/2} K_\mu(2\sqrt{\beta\gamma}), \qquad \Re(\beta)>0,\ \Re(\gamma)>0,
\label{eq:B11} \tag{B.11}
\end{equation}
with $\mu=2\nu+1+m+m'$, $\beta=2c/N_\nu$ and $\gamma=1$, we obtain
\begin{equation}
I^{\mathrm{HF}}_{n_r,l} = 2\,\Gamma(n_r+2\nu+2)^2 \sum_{m=0}^{n_r}\sum_{m'=0}^{n_r} \chi^{(n_r,\nu)}_{mm'} \left(\frac{2c}{N_\nu}\right)^{(2\nu+1+m+m')/2} K_{2\nu+1+m+m'}\!\left(2\sqrt{\frac{2c}{N_\nu}}\right).
\label{eq:B12} \tag{B.12}
\end{equation}

Substituting eq.~\ref{eq:B12} into eq.~\ref{eq:B5}, we arrive at the general Hellmann-Feynman integrand
\begin{align}
\left\langle \frac{e^{-c/r}}{r^2} \right\rangle^{\mathrm{Kr}}_{n_r,l} &= \frac{4\,\Gamma(n_r+1)\,\Gamma(n_r+2\nu+2)}{N_\nu^3} \nonumber \notag \\
& \sum_{m=0}^{n_r}\sum_{m'=0}^{n_r} \chi^{(n_r,\nu)}_{mm'} \left(\frac{2c}{N_\nu}\right)^{(2\nu+1+m+m')/2} K_{2\nu+1+m+m'}\!\left(2\sqrt{\frac{2c}{N_\nu}}\right)
\label{eq:B13} \tag{B.13}
\end{align}
with $\chi^{(n_r,\nu)}_{mm'}$ given by eq.~\ref{eq:B9}.

For the ground state $(n_r=l=0)$, only the term $m=m'=0$ contributes, so eq.~\ref{eq:B13} reduces to
\begin{equation}
\left\langle \frac{e^{-c/r}}{r^2} \right\rangle^{\mathrm{Kr}}_{1s} = \frac{4}{N_\nu^3\,\Gamma(2\nu+2)} \left(\frac{2c}{N_\nu}\right)^{\nu+1/2} K_{2\nu+1}\!\left(2\sqrt{\frac{2c}{N_\nu}}\right)
\label{Eq:B14} \tag{B.14}
\end{equation}
in agreement with the expression quoted in the main text.

The Hellmann-Feynman energy is then obtained from
\begin{equation}
E_{n_r,l}(c) = -\frac{1}{2n^2} + \int_0^c \left\langle \frac{e^{-c'/r}}{r^2} \right\rangle^{\mathrm{Kr}}_{n_r,l}(c')\,dc', \qquad n=n_r+l+1.
\label{eq:B15} \tag{B.15}
\end{equation}
Since both $\nu_l(c')$ and $N_\nu(c')$ depend on the integration variable $c'$, the full Hellmann-Feynman Kratzer energy is evaluated numerically.

%=============================================================================
\section*{Acknowledgments}
%=============================================================================

This work was supported by the Algerian Ministry of Higher Education and Scientific Research PRFU B00L02UN050120230005.

%=============================================================================

\end{document}